\documentclass[11pt,a4paper]{article}
\usepackage[english]{babel}
\usepackage{amsmath,amsthm,amssymb,epsfig,latexsym}
\usepackage[colorlinks=true,linkcolor=blue,citecolor=magenta]{hyperref}
\usepackage{color}
\usepackage{ulem}
\usepackage{cancel}
\usepackage{bm}
\usepackage[numbers,square,sort&compress]{natbib}
\def\fg{\mathfrak{g}}

\def\fl{\mathfrak{l}}

\def\cV{\mathcal{V}}
\def\gl#1{\fg\fl_{#1}}

\newcommand{\so}{{\scriptscriptstyle \rm I}}
\newcommand{\st}{{\scriptscriptstyle \rm I\hspace{-1pt}I}}
\newcommand{\sth}{{\scriptscriptstyle \rm I\hspace{-1pt}I\hspace{-1pt}I}}

\newcommand{\bu}{\bar u}
\newcommand{\bv}{\bar v}
\newcommand{\bx}{\bar x}
\newcommand{\by}{\bar y}
\newcommand{\bz}{\bar z}
\newcommand{\bt}{\bar t}
\newcommand{\bw}{\bar w}
\newcommand{\be}[1]{\begin{equation}\label{#1}}
\newcommand{\ba}[1]{\begin{multline}\label{#1}}
\newcommand{\ee}{\end{equation}}
\newcommand{\ea}{\end{multline}}

\newcommand{\tr}{\mathop{\rm tr}}

\newtheorem{thm}{Theorem}[section]
\newtheorem{prop}[thm]{Proposition}
\newtheorem{lemma}[thm]{Lemma}
\newtheorem{rem}[thm]{Remark}
\newtheorem{cor}[thm]{Corollary}
\newtheorem{Def}[thm]{Definition}

\def\qed{\hfill\nobreak\hbox{$\square$}\par\medbreak}
 \makeatletter
 \@addtoreset{equation}{section}
 \makeatother
 
\newcommand{\bea}{\begin{eqnarray}}
\newcommand{\eea}{\end{eqnarray}}

\newcommand{\bleu}[1]{{{\color{blue} #1}}}

\def\BB{{\mathbb{B}}}

\def\CC{{\mathbb{C}}}
\newcommand{\ZZ}{{\mathbb Z}}
\def\TT{{\mathbb{T}}}

\def\rvac{|0\rangle}

\def\TT{\mathsf{T}}

\def\Ee{{\sf e}}

\def\r#1{\eqref{#1}}

\def\ggo{\mathfrak{g}}
\def\hgo{\mathfrak{h}}

\def\ot{\otimes}

\def\Xif#1#2#3#4#5#6#7#8
{\Xi^{0,1}_{#1,#2}
{\small\Big(\begin{array}{ccc}#3&#4&#5\\ #6&#7&#8 \end{array}\Big)}}

\def\Oml{\Omega^{L}}
\def\Omr{\Omega^{R}}

\def\Z{\mathcal{Z}}
\def\vn{\varnothing}

\def\eigen{\tau}
\def\Om{\Omega}

\def\Ups{\Upsilon}

\def\crd#1#2{\mathop{#1}\limits_{#2}}

\def\phf{{\Lambda}}
\def\coA{{\Gamma}}
\def\roA{\overline{{\Gamma}}}

\def\bgam{\bm{\gamma}}

\def\nus{\nu}

\def\Ig{I_{\mathfrak{g}}}
\def\rvec{|0\rangle}

\def\LL{T}
\def\Alm{\mathfrak{A}}
\def\rt{\mathsf{r}}
\def\Tzm{\mathfrak{T}}

\def\ophf{\overline{{\Lambda}}}

\def\sPh{\Phi}
\def\oPh{\overline{\Phi}}
\def\PSR{J}
\def\frd{\xi}

\def\bvphi{\varphi}
\def\gln{\mathfrak{gl}_n}
\def\ggb{\mathfrak{o}_{2n+1}}
\def\ggd{\mathfrak{o}_{2n}}
\def\gsp{\mathfrak{sp}_{2n}}

\def\col{\mathfrak{q}}
\def\Sk{\mathfrak{k}}
\def\fd{\mathfrak{d}}
\def\bup{\bar{\vartheta}}
\def\bvp{\bar{\vartheta}}

\begin{document}

\vspace{12pt}

\begin{center}
\begin{LARGE}
{\bf Bethe Vectors in Quantum Integrable Models\\[3mm] with Classical Symmetries}
\end{LARGE}

\vspace{10mm}

\begin{large}
A.~Liashyk${}^{a}$, S.~Pakuliak${}^{b}$ and E.~Ragoucy${}^{b}$
\end{large}

\vspace{10mm}

${}^a$ {\it Beijing Institute of Mathematical Sciences and Applications (BIMSA),\\
No. 544, Hefangkou Village Huaibei Town, Huairou District Beijing 101408, China}

\vspace{2mm}

${}^b$ {\it Laboratoire d'Annecy de Physique Théorique(LAPTh)\\ 
CNRS $\&$ Universit\'e Savoie Mont Blanc\\
Chemin de Bellevue, BP 110, F-74941, Annecy-le-Vieux cedex, France}
\vspace{2mm}

E-mails: liashyk@bimsa.cn, pakuliak@lapth.cnrs.fr, ragoucy@lapth.cnrs.fr


\end{center}

\begin{center}

\end{center}

\begin{abstract}
\noindent 
The first goal of this paper is to give a precise and simple definition for off-shell Bethe vectors in a generic $\ggo$-invariant integrable model for 
$\ggo=\mathfrak{gl}_n$, $\mathfrak{o}_{2n+1}$, $\mathfrak{sp}_{2n}$ and 
$\mathfrak{o}_{2n}$.
We prove from our definition that the off-shell Bethe vectors indeed become on-shell when the Bethe equations are obeyed. 

Then, we show that some properties for these off-shell Bethe vectors, such as the action 
formulas of monodromy entries on these vectors, their
rectangular recurrence relations and their coproduct formula, are a consequence of our definition.
\end{abstract}


\section{Introduction}

The algebraic Bethe ansatz is a very powerful tool to obtain the eigenvectors of the transfer matrix defining a quantum integrable spin chain. In this context, these eigenvectors are called \textsl{on-shell Bethe vectors}. The term \textsl{on-shell} refers to the relations (the Bethe equations) that the parameters (the Bethe parameters) entering the Bethe vectors have to satisfy, in order to be true eigenvectors of the transfer matrix. Then, it is natural to call \textsl{off-shell} the same Bethe vectors defined without assuming that their Bethe parameters obey the Bethe equations. In other words, 
 off-shell Bethe vectors are loosely defined as becoming on-shell when the Bethe equations are obeyed. 
 This definition suffers from a lack of precision, since any term which vanishes when the Bethe equations are fulfilled can be added to a Bethe vector without spoiling this loose definition.
 However, off-shell Bethe vectors play an important role in the calculation of form factors and correlation functions, so that a precise definition of these objects is essential. 
The first goal of this paper is to give a precise definition for off-shell Bethe vectors
 and prove that they indeed become on-shell when the Bethe equations are obeyed. The second goal is to show that some properties for these off-shell Bethe vectors, sometimes used as definitions for them, are a consequence of our definition.

The framework used in this paper follows the one developped in 
the papers  \cite{HLPRS20,LP2,LPR25}.
 In these papers, the
algebraic Bethe ansatz for a generic $\ggo$-invariant integrable 
model was developed for the finite-dimensional Lie algebras 
$\mathfrak{gl}(m|n)$ and $\mathfrak{o}_{2n+1}$. By a generic 
$\ggo$-invariant integrable model we understand a model which 
is defined by a monodromy matrix $T(u)$ satisfying the $RTT$ commutation 
relations  with a $\ggo$-invariant $R$-matrix, with no assumption on the representation space $T(u)$ acts on (except that it is highest weight). For Yangians based on finite-dimensional 
algebras corresponding to the classical series, these $R$-matrices 
were discovered long ago in \cite{Yang,ZZ79}. The $RTT$ commutation 
relations are the defining relations for the Yangian $Y(\ggo)$ in its 
matrix realization \cite{Drinfeld,Molev}.   Representations of the 
Yangians $Y(\ggo)$ using $\ggo$-invariant 
$R$-matrices were presented in  \cite{AMR}. Isomorphisms between 
$R$-matrix and Drinfeld's realizations of 
$Y(\ggo)$ was done 
in \cite{DF, JLM18}. It allows to replace the monodromy 
matrix in a generic $\ggo$-invariant model by the fundamental 
matrix operator $T(u)$ 
of the Yangian $Y(\ggo)$.

Due to the $RTT$ commutation relation the trace of the 
fundamental monodromy matrix is a generating series for 
commuting operators and the main objective of the algebraic 
Bethe ansatz is to construct
the eigenvectors of these commuting operators in terms of the 
monodromy entries. 
Combinatorial expressions for these eigenvectors can be found in 
the framework of nested Bethe ansatz as certain polynomials of non-commuting monodromy entries 
acting on a vacuum vector for $\mathfrak{gl}_N$\cite{KR1,KR2} and for $\mathfrak{o}_{N}$, $\mathfrak{sp}_{2n}$ \cite{DVK, Res85, Res91}.
Another possibility is to use a trace formula for the Bethe vectors found in [28] for $\mathfrak{gl}_n$ and $U_q(\mathfrak{gl}_n)$. Later,  the trace formula was found for the $C$ and $D$ seria cases and their $q$-deformed generalizations, see \cite{GR}.
Unfortunately, the trace formula is not known for $B$-seria case.

An alternative approach to construct Bethe vectors was proposed in [8, 14] for the integrable models related to $U_q(\widehat{\mathfrak{gl}}_N)$.
It was called the \textsl{projection method} because it uses  
projections to the intersections of different type Borel subalgebras in 
the quantum affine algebras acting on the product of simple 
root currents in the Drinfeld's 'new' realization \cite{Dnew}
of the quantum affine algebras
$U_q(\widehat{\ggo})$. 
This approach was generalized for the Yangian double $DY(\mathfrak{gl}(m|n)$ in \cite{HLPRS17} and
for $DY(\mathfrak{o}_{2n+1})$ in \cite{LP2}.
It allows to construct the so-called off-shell Bethe vectors, i.e. when their Bethe parameters are free.
The off-shell Bethe vectors 
becomes on-shell Bethe vectors and diagonalize the commuting set 
of integrals when their parameters satisfy the Bethe equations. 

The paper \cite{HLPRS17} presents the action of the upper-diagonal and diagonal 
monodromy entries on the off-shell Bethe vectors. 
Nested recurrence relations corresponding to the extreme nodes of the 
Dynkin diagram were also obtained in this paper
in the framework of the projection method. 
Unfortunately, the projection method does not allow 
to get easily the action formula of an arbitrary monodromy entry,
nor the recurrence relations corresponding to an arbitrary node of the 
Dynkin diagram. 
To get these actions, another method, called the zero mode method, was designed in \cite{HLPRS20} for $\mathfrak{gl}(m|n)$-invariant 
models. It was then generalized in \cite{LPR25} to get recurrence relations 
corresponding to an
arbitrary node of the Dynkin diagram for the off-shell Bethe vector of
$\mathfrak{gl}_n$ and $\mathfrak{o}_{2n+1}$-invariant models. The calculation was done inductively with the starting 
point taken from the projection method.

It was shown in the papers \cite{HLPRS20} and \cite{LPR25}
that when $\ggo=\mathfrak{gl}_n$ or 
$\mathfrak{o}_{2n+1}$, in order to fully describe off-shell and on-shell Bethe vectors 
in terms of the action of an arbitrary monodromy entry and 
the recurrence relations with respect to an arbitrary node of the Dynkin diagram
one has to combine a single (rather simple) action of the most upper right entry of the monodromy matrix (called below the highest monodromy entry) with the action of  certain zero modes of monodromy matrix
identified with the simple root generators of the algebra $\ggo$.  
This data can be obtained from elsewhere, in particular from the 
projection method.  Once this data obtained (or postulated), the
full description of the Bethe vectors in terms of the arbitrary 
monodromy entry the action formulas and the recurrence relation 
corresponding to an arbitrary Dynkin node is done using basically 
only $RTT$ commutation relations.  This opens the possibility to define off-shell 
Bethe vectors without any reference to the Drinfeld's 'new' realization 
 and the use of the Yangian doubles.

In this paper we realize this program for arbitrary algebra Lie $\ggo$ 
of the classical series. We start to consider a generic 
$\mathfrak{gl}_n$-invariant model to recall some already known results recasted within our approach.
Then we repeat this program for the algebras $\mathfrak{o}_{2n+1}$,  
$\mathfrak{sp}_{2n}$, and $\mathfrak{o}_{2n}$. 
Surprisingly enough, the formulas for 
the monodromy entries actions and the recurrence relations 
can be unified and written as two formulas valid for any generic 
$\ggo$-invariant integrable model. It is achieved by
introducing two discrete parameters $\epsilon_\ggo$ and $\xi_\ggo$, whose algebraic interpretation remains to be clarified.  

We would also like to emphasize that the results presented in this article are \textsl{universal}, in the sense that they apply to any integrable spin-chain model, regardless of the representations used at each site (provided they are highest-weight representations). As such, all the functions appearing in the various formulas are the same for any model. The only model dependence is reflected in the functional parameters
$\lambda_j(z)$ and in their ratios $\alpha_s(z)$.

This paper is organized as follow. 
In section~\ref{sect2} we shortly present the main results 
of this paper. Section~\ref{sect3} is devoted to 
the description of generic $\ggo$-invariant models.
The properties of off-shell
and on-shell Bethe vectors are described in full 
details for $\ggo=\gln$ in the 
section~\ref{sect4} and 
for algebras $\ggo=\mathfrak{o}_N$ and $\mathfrak{sp}_{2n}$ in the section~\ref{sect5}.
The section~\ref{sect6} is devoted to the coproduct properties 
of off-shell Bethe vectors resulting from their defining relations.
Appendix~\ref{AppB} contains definitions of 
the rational functions used in the core of the paper 
and appendix~\ref{AppA} describes the cardinalities 
of the subsets in the action of the monodromy diagonal
entries on the off-shell Bethe vectors.

\section{Main results}\label{sect2}
The construction we will present is rather technical and we summarize here the main results of our paper, namely a precise definition of off-shell Bethe vectors, valid for any Yangian based on the $A$, $B$, $C$, and $D$ series, together with the properties which come out from this definition.
Since this section is a summary, we will not give details, and focus on the general features, common to all the algebras mentioned above. We refer to the core of the paper for a more precise description. 

We start with a $RTT$ presentation for the Yangian $Y(\fg)$, $\fg$ being one of the Lie algebras $\mathfrak{gl}_n$, $\mathfrak{o}_{2n+1}$, $\mathfrak{sp}_{2n}$ and $\mathfrak{o}_{2n}$. The framework is precisely defined  in the next section, here we just assume that we have a monodromy matrix $T(z)$ whose entries $T_{i,j}(z)$ are labeled by indices running over a set $I_\fg$ which depends on the algebra we consider. The cardinality of $I_\fg$ is $N_\fg$ and corresponds to the dimension of the fundamental representation of $\fg$. The set $I_\fg$ is invariant under an idempotent mapping $i\to i'$ which depends on $\fg$, see next section. Using this mapping, we can write $I_\fg=\{n',n'+1,...,n-1,n\}$. $I_\fg$ contains a subset $J_\fg$ which is used to label the simple roots of $\fg$. Clearly the cardinality of $J_\fg$ is the rank of $\fg$. We can write $I_\fg=J_\fg\cup(J_\fg)'\cup\{n',n\}$, noting that the sets $J_\fg$, $(J_\fg)'$ and $\{n',n\}$ may have non-empty intersections.

The \textsl{zero modes} $\TT_{i,j}$ of $T_{i,j}(z)$ define a subalgebra $\fg$ of $Y(\fg)$ and we will use its Cartan subalgebra to provide \textsl{colors} $a\in J_\fg$ to the monodromy matrix entries and to formal parameters. We will note $\col_a$, $a\in J_\fg$, the Cartan generators whose eigenvalues correspond to the colors, and $\Tzm_a$, $a\in J_\fg$, the generators associated to the simple roots of $\fg$.

We consider generalized models, defined by a vacuum $|0\rangle$, a set of functions $\lambda_i(u)$, $i\in I_{\fg}$ and the relations
$\big(T_{ij}(u)-\delta_{ij}\lambda_i(u)\big)|0\rangle=0$, $i\geq j$, valid for any formal parameter $u$. It is known that all spin chain models fall in this class of models, see for instance \cite{LPR25}. 
The transfer matrix of these models is $\tr\Big(T(z)\Big)$, which commutes with the algebra $\fg$. Then, we can look for eigenvectors of both the transfer matrix and of the color generators $\col_a$, $a\in J_\fg$.
The eigenvectors of these models belong to the space $\cV^+$ of polynomials in the non-commutative elements $T_{i_1,j_1}(u_1), \ldots, T_{i_k,j_k}(u_k)$, $i_\ell<j_\ell$, $1\le \ell \le k$, for formal parameters $u_k$, ${k\in\ZZ_{> 0}}$, and acting on the vacuum.

\begin{Def}\label{def:BV}
Let $\bar t$ be a collection of sets $\bar t=\{\bar t^a,\, a\in J_\fg\}$, where each set $\bar t^a$ gathers  formal parameters of color $a$.
The off-shell Bethe vectors $\BB(\bar t)$ are element of $\cV^+$, uniquely defined (up to a normalisation factor) by the initial condition $\BB(\vn)=|0\rangle$ and the following relations
\begin{equation*}\label{defBV}
\begin{split}
T_{n',n}(z)\, \BB(\bar t) &=\lambda_n(z)\,\mu^n_{n'}(z;\bar t)\,\BB(\bw), \\
 \Tzm_a\, \BB(\bar t) &= 
 \sum_{{\rm part}} \Big(\alpha_a(\bar t^a_{\so})\, \Oml_a(\bt_{\st}, \bt_{\so}) - \Omr_a(\bt_{\so}, \bt_{\st})\Big)\, \BB(\bar t_{\st}), \quad a\in J_\ggo,
\end{split}
\end{equation*}
where $\mu^n_{n'}(z,\bar t)$, $\Oml_a(\bar t_{\st}, \bar t_{\so})$ and $\Omr_a(\bar t_{\so}, \bar t_{\st})$ are rational functions, precisely defined (for a specific normalization of the off-shell Bethe vectors) in equations \eqref{T-act-gl}, \eqref{T-act-g}, \eqref{zm-ac2} and \eqref{bvde2}. The sum runs over partitions $\bar t^s\dashv \{\bar t^s_{\so},\bar t^s_{\st}\}$ with $|\bar t^s_{\so}|=\delta_{a,s}$. The set $\bar w$ is an extension of $\bar t$ by $z$, see its precise definition (which depends on $\fg$) in \eqref{exA}, \eqref{z-sh2} and \eqref{exBCD}. The functions $\alpha_a(z)$ are ratio of 
$\lambda_i(z)$ functions, see \eqref{alpha-gl} and \eqref{alpha-all}.
\end{Def}

To present the different properties associated to this definition, we need to introduce the operators 
${\mathcal{Z}}^k_l$ which extend the sets $\bar t^a$ by the formal parameter $z$ and/or a shifted version of $z$. The precise definition of ${\mathcal{Z}}^k_l$ and of the shifts of $z$ depend on the algebra $\fg$, and can be found in equations \eqref{Zop-gl}, \eqref{z-sh1} and \eqref{Zop-all}. 

The normalisation factor used in the definition \ref{def:BV} is consistent with the functions described below.

Then, the unicity of the off-shell Bethe vectors is proved thanks to the following property, since it
allows to build the Bethe vectors starting from the initial condition.
\begin{prop}\label{an-rrr}
Let $\BB(\bar t)$ be an off-shell Bethe vector. Then, it obeys the following rectangular recurrence relations
\begin{equation*}
\Z^k_\ell\cdot\BB(\bt)=\frac{1}{ \lambda_k(z)\,\mu^k_\ell(z;\bt)}\ 
\sum_{i=n'}^\ell\sum_{j=k}^n\sum_{{\rm part}}
\alpha(\bt_\sth)\,
\Xi^{\ell,k}_{i,j}(z;\bt_{\so},\bt_{\st},\bt_{\sth})\,T_{i,j}(z)\cdot\BB(\bt_{\st})\,,
\end{equation*}
where $\mu^k_{\ell}(z;\bar t)$ and $\Xi^{\ell,k}_{i,j}(z;\bt_{\so},\bt_{\st},\bt_{\sth})$ are rational functions, precisely defined in equations \eqref{p-p-m-gl}, \eqref{mu-all}, \eqref{Xi-rr-gln} and \eqref{Xi-rr}.
The sum runs over partitions $\bar t^a\dashv \{\bar t^a_{\so},\bar t^a_{\st},\bar t^a_{\sth}\}$ with cardinalities of $\bar t^a_{\so}$ and $\bar t^a_{\sth}$ depending on the algebra $\fg$, and the indices $i,j,k,\ell$, see \eqref{oCgl} and \eqref{oC-all}.
\end{prop}

In \cite{LPR25}, the recurrence relations of the proposition~\ref{an-rrr} were called {\sl rectangular} because their r.h.s. contains  monodromy matrix elements from 
a rectangular domain of the monodromy matrix, see section \ref{sec:rec-rec}.

The second property can be viewed as a dual version of the previous one since it can be used to
prove the recurrence relations.
\begin{prop}
Let $\BB(\bar t)$ be an off-shell Bethe vector. Then, the action of the monodromy matrix entries reads
\begin{equation*}
T_{i,j}(z)\cdot \BB(\bt)=\lambda_n(z)\,\mu^n_{n'}(z;\bt)\,
\sum_{{\rm part}\ \bw^s}
\BB(\bw_{\st})\, \frac{\coA_i(\bw_{\so})}{\psi_{n'}(z;\bw_{\so})}\, 
\frac{\roA_j(\bw_{\sth})}{\phi_n(z;\bw_{\sth})}\, 
\Om(\bw_{\so},\bw_{\st},\bw_{\sth})\,  
\alpha(\bw_{\sth})\,,
\end{equation*}
where  $\coA_i(\bw_{\so})$, $\psi_{n'}(z;\bw_{\so})$,
$\roA_j(\bw_{\sth})$ and $\phi_n(z;\bw_{\sth})$ are rational functions, see propositions \ref{ac-gln} and \ref{ac-main}.
\end{prop}

\begin{rem}
We note that the definition of off-shell Bethe vectors is very similar to the construction of the  root system  for affine algebras starting from the simple roots of the finite algebra  $\ggo$. Indeed, the zero modes $\Tzm_a$ (located in the lower triangular part of the monodromy matrix) can be clearly related to the simple roots $\rt_a$ of the corresponding finite dimensional algebra $\ggo$, while the highest entry $T_{n',n}(z)$ (located in the upper triangular part of the monodromy matrix)  is reminiscent of the simple affine root $-\sum_{a\in \PSR_\ggo}\rt_a+\delta$.  Then using this root and the simple roots
of $\ggo$ one  builds half part of the whole affine root system, just like we reconstruct the action of all $T_{i,j}(u)$ from the action of the zero modes $\Tzm_a$ and the highest monodromy $T_{n',n}(z)$. 
\end{rem}

\begin{cor}
The off-shell Bethe vectors have definite colors:
\begin{equation*}
\col_{a}\, \BB(\bar t) = |\bar t^a|\, \BB(\bar t)\,, \quad a\in J_\ggo\,.
\end{equation*}
\end{cor}

The next property justifies the name off-shell Bethe vectors we use in the definition \ref{def:BV}:
\begin{prop}
Let $\BB(\bar t)$ be an off-shell Bethe vector. If the Bethe equations
\begin{equation*}
\alpha_a(\bar t^a_{\so})=\frac{\Omr_a(\bt_{\so},\bt_{\st})}{\Oml_a(\bt_{\st},\bt_{\so})}\,,
\qquad a\in J_\fg
\end{equation*}
are obeyed, $\BB(\bar t)$ is an eigenvector of the transfer matrix $\mathcal{T}_{\ggo}(z)=\tr T(z)$:
\begin{equation*}
\mathcal{T}_\ggo(z)\,\BB(\bar t) = \tau_\ggo(z;\bar t)\,\BB(\bar t)\,.
\end{equation*}
Moreover, the on-shell Bethe vectors are also highest weight vectors for the $\fg$ algebra:
\begin{equation*}
    \Tzm_a\, \BB(\bar t) =0\,,\quad a\in J_\fg\,.
\end{equation*}
The explicit form of the eigenvalues are given in \eqref{BEA9}, \eqref{BEB3}, \eqref{BEC3} and \eqref{BED2}.
\end{prop}

The next proposition allows to build Bethe vectors for composite models.
It is usually called the coproduct formula for off-shell Bethe vectors.
 \begin{prop}
Let $\BB^{[2]}(\bt)$ and $\BB^{[1]}(\bt)$ be two off-shell Bethe 
 vectors obeying the definition~\ref{def:BV} for two different commuting monodromy matrices $T^{[2]}(u)$ and $T^{[1]}(u)$ respectively. Then, 
 the composed Bethe vector 
 \begin{equation*}
 \BB^{[2,1]}(\bt)=\sum_{\rm part}\Om(\bt_\so,\bt_\st)\,
 \BB^{[2]}(\bt_\so)\,\BB^{[1]}(\bt_\st)\ \prod_{s\in\PSR_\ggo}
 \alpha_s^{[2]}(\bt^s_\st)\,
 \end{equation*}
 also obeys the definition \ref{def:BV} for the composite model defined by the monodromy matrix
 \begin{equation*}
 T^{[2,1]}(z)=T^{[2]}(u)\,T^{[1]}(u),\quad\mbox{i.e.}\quad
 T_{i,j}^{[2,1]}(z)=\sum_{k=n'}^n\, T^{[2]}_{i,k}(u)\,T^{[1]}_{k,j}(u)\,.
 \end{equation*}
 The rational function $\Omega$ is defined in \eqref{Om-gen}.
 \end{prop}

To end this section, we would like to stress that the above definition is directly inherited from the current presentation of the double Yangian $DY(\fg)$. Indeed, an alternative presentation of off-shell Bethe vectors (called universal weight functions in this context) has been introduced using the current presentation and a projection on different types of Borel subalgebras of $DY(\fg)$ \cite{LPR-PM}. In this framework, the properties described above (including the formulas used in the definition \ref{def:BV}) has been proven for $\mathfrak{gl}_n$ and $\mathfrak{o}_{2n+1}$ algebras, and the correspondence with a "usual" definition of Bethe vectors (using the RTT presentation of $Y(\fg)$) was done on a case-by-case basis, \cite{HLPRS20,LP2}.
In the present paper, the proofs rely only on the $RTT$ presentation, thereby establishing the full compatibility between these two definitions.

\section{Generic $\ggo$-invariant model}\label{sect3}

Let $\mathfrak{g}$ be one of the classical Lie algebras $\mathfrak{gl}_n$, $\mathfrak{o}_{2n+1}$, 
$\mathfrak{sp}_{2n}$ and $\mathfrak{o}_{2n}$ corresponding to the $A_{n-1}$, 
$B_n$, $C_n$ and $D_n$ classical series.
Note that the algebra $\mathfrak{gl}_n$ will be sometimes singled out in this list, because it is the only one that can be embedded in all other algebras. We will heavily use this property to shorten some proofs. It explains why the list is $A_{n-1}$, $B_{n}$, $C_{n}$, $D_{n}$ despite the rank of $A_{n-1}$ is different from the other ones.

\subsection{$R$-matrix}
Fix a positive integer $n\geq2$. 
Let $V_\ggo$ be a vector space for the fundamental representation of $\ggo$. 
We note $N_\ggo$ the dimension of $V_\ggo$. 
For the classical Lie algebras $\ggo=\mathfrak{gl}_n$, $\mathfrak{o}_{2n+1}$, 
$\mathfrak{sp}_{2n}$, $\mathfrak{o}_{2n}$, this dimension
 is $n$, $2n+1$, $2n$ and $2n$  respectively.
In order to describe the operators (matrices) acting in these 
representations for different $\ggo$ on the equal footing we introduce following  
notations. 
Let $\frd_\ggo$ be the integer parameter such that 
\begin{equation*}\label{N-g}
N_\ggo\equiv N=2n+1-\frd_\ggo\,.
\end{equation*} 
Explicitly
\begin{equation}\label{prime-par}
\frd_\ggo=\begin{cases}
n+1,&\mathfrak{g}=\mathfrak{gl}_{n}\,,\\
0,&\mathfrak{g}=\mathfrak{o}_{2n+1}\,,\\
1,&\mathfrak{g}=\mathfrak{sp}_{2n},\ \mathfrak{o}_{2n}\,.
\end{cases}
\end{equation}

For any integer $i$ we define a map $i\to i^\prime$ such that 
\begin{equation}\label{prime}
i^{\,\prime}=\frd_\ggo-i\,.
\end{equation}
This map leaves invariant the set of matrix indices 
\begin{equation*}\label{I-g}
\Ig=\{n',n'+1,\ldots,n-1,n\}\,
\end{equation*}
of operators from ${\rm End}(V_\ggo)$.

Let $e_a$ be an orthonormal basis 
\begin{equation*}\label{ortbas}
(e_a,e_b)=\delta_{a,b}\, ,\qquad a,b=1 ,\ldots,N
\end{equation*}
with respect to the natural scalar product in $\CC^N$.

Let $\PSR_{\mathfrak{g}}$ be the set of indices corresponding to the nodes of the
 $\mathfrak{g}$-type Dynkin diagram:
\begin{equation*}\label{PSR-g}
\PSR_{\mathfrak{g}}=
\begin{cases}
\{1,2,\ldots,n-1\}\,, \qquad  & \mathfrak{g} = \mathfrak{gl}_n ,\\     
\{0,1,2,\ldots,n-1\}\,, & \mathfrak{g} = \mathfrak{o}_{2n+1},\ \mathfrak{sp}_{2n}, \  \mathfrak{o}_{2n}.
\end{cases}
\end{equation*}

The positive simple roots $\rt_a$, $a\in\PSR_\ggo$ for the algebra $\ggo=\mathfrak{gl}_n$ are 
\begin{equation*}\label{sim-r-g}
\begin{aligned}
 &&&\rt_{a}=e_{a+1}-e_{a}\,, &&a=1,..,n-1,   
 &&\ggo=\mathfrak{gl}_{n}\,, \\
\rt_0&=e_1, &&\rt_a=e_{a+1}-e_a, &&a=1,..,n-1, && 
\ggo=\mathfrak{o}_{2n+1}\,,\\
\rt_0&=2\,e_1, &&\rt_a=e_{a+1}-e_a, &&a=1,..,n-1, 
&&\ggo=\mathfrak{sp}_{2n}\,,\\
\rt_0&=e_2+e_1,\ &&\rt_a=e_{a+1}-e_a,\ &&a=1,..,n-1,   
\quad &&\ggo=\mathfrak{o}_{2n}\,.
\end{aligned}
\end{equation*}
Let $\hgo_{a,b}$  be the symmetric matrix of the simple root scalar products $\hgo_{a,b}=(\rt_a,\rt_b)$. In the case $\mathfrak{g} = \mathfrak{gl}_n$
\begin{equation*}\label{Cmat-gl-n}
 \hgo_{a,b}=(\rt_a,\rt_b)=2\delta_{a,b}-\delta_{a,b-1}-\delta_{a,b+1}\, \qquad \qquad 1\leq a,b\leq n-1, 
\end{equation*}
and for $\mathfrak{g} = \mathfrak{o}_{2n+1},\ \mathfrak{sp}_{2n}, \  \mathfrak{o}_{2n}$
\begin{equation*}
\begin{aligned}
 \hgo_{a,b}&=2\,\delta_{a,b}-\delta_{a,b-1}-\delta_{a,b+1}\,,\qquad\qquad\qquad\qquad &&2\leq a,b\leq n-1\,,\\
 \hgo_{a,1}&=\hgo_{1,a}=2\,\delta_{a,1}-\delta_{a,2}-(2-\xi_\ggo-\epsilon_\ggo)\,\delta_{a,0}\,, 
\ \quad &&0\leq a\leq n-1\,,\\
 \hgo_{a,0}&=\hgo_{0,a}=(2+\xi_\ggo-\epsilon_\ggo)\,\delta_{a,0}-
(3-\epsilon_\ggo)\,\delta_{a,2}/2
\,,  \quad &&0\leq a\leq n-1\,.
\end{aligned}
\end{equation*}
where we introduced the parameter $\epsilon_{\ggo}$:
\begin{equation}\label{eps-g}
\epsilon_{\ggo}=\begin{cases} 
0, &\mbox{ for } \ggo=\mathfrak{gl}_n\,, \\
1, &\mbox{ for } \ggo=\mathfrak{o}_N\,, \\
-1, &\mbox{ for } \ggo=\mathfrak{sp}_{2n} \,,
\end{cases}
\end{equation}
 As we shall see below, the same parameter $\epsilon_{\ggo}$ allows to define the $R$-matrix in a uniform way.

Accordingly, let us introduce and "local" versions of $\epsilon_{\ggo}$:
\begin{equation}\label{eps-i}
\epsilon_{i}=(\epsilon_{\ggo})^{\delta_{i>0}}\,,\quad i\in I_\ggo\,,
\end{equation}
where $\delta_{\rm condition}$ is equal to 1 when the "condition" is satisfied and 
equal to zero otherwise. These parameters allows to define the operation $(\cdot)^{\rm t}$ acting on $N \times N$ square matrices $M$ as
\begin{equation}
\big(M^{\rm t}\big)_{i,j}=\epsilon_i\,\epsilon_j\,M_{j'\!,i'}\,,\quad i,j\in I_\ggo\,.
\end{equation}
When $\ggo\neq \gln$, $(\cdot)^{\rm t}$ is a transposition, but it gives 0 for $\ggo= \gln$. We will nevertheless loosely call it a transposition. 
With the help of this tranposition the $\mathfrak{g}$-invariant $R$-matrix  \cite{Yang,ZZ79} reads
\begin{equation}\label{RABCD}
  R(u,v) = \mathbf{I}\otimes\mathbf{I} + \frac{c\, \mathbf{P}}{u-v} - \frac{c\,\mathbf{Q}}{u-v+c\,\kappa_{\mathfrak{g}}}\,,
\end{equation}
where $\mathbf{I}=\sum_{i\in\Ig}\Ee_{i,i}$ is the identity operator acting in the space $\CC^{N}$ and
$\Ee_{i,j}\in{\rm End}(\CC^{N})$ are $N\times N$ 
matrices with the only nonzero entry equals to 1 at 
the intersection of the $i$-th row and $j$-th column. The operators $\mathbf{P}$ 
and $\mathbf{Q}=\mathbf{P}^{{\rm t}_2}=\mathbf{P}^{{\rm t}_1}$ act in 
$\CC^{N}\ot\CC^{N}$ as
\begin{equation*}\label{PQ}
\mathbf{P}=\sum_{i,j\in \Ig}\Ee_{i,j}\otimes\Ee_{j,i}, \quad 
\mathbf{Q}=\sum_{i,j\in \Ig}\epsilon_i\,\epsilon_j\,\Ee_{i,j}\otimes\Ee_{i'\!,j'}\,.
\end{equation*}
Note that $\mathbf{Q}=0$ for $\ggo=\gln$.
The $R$-matrix \r{RABCD} depends on the formal spectral parameters 
$u$, $v$ and the rational functions of these parameters entering the
definition of $R(u,v)$ should be understood as Laurent series with respect 
to either the ratio $u/v$ or the ratio $v/u$. 

The parameter $\kappa_\ggo$ entering the definition of the $\ggo$-invariant 
$R$-matrix \r{RABCD} needs to be defined only for $\ggo\neq\gln$. It reads
\begin{equation*}\label{kappa}
\kappa_\ggo=n+\frac12(1-\xi_\ggo)-\epsilon_{\ggo}\equiv n+\theta_\ggo\,,
\end{equation*}
where to lighten the presentation, we have introduced
\begin{equation}\label{t-vp}
\theta_{\mathfrak{g}}= \frac12(1-\xi_\ggo)-\epsilon_{\ggo}=
\begin{cases}
-1/2,&\mathfrak{g}=\mathfrak{o}_{2n+1}\,,\\
1,&\mathfrak{g}=\mathfrak{sp}_{2n}\,,\\
-1,&\mathfrak{g}=\mathfrak{o}_{2n}\,.
\end{cases}
\end{equation}
We will also use
\begin{equation}\label{vphi}
\bvphi_{\mathfrak{g}}=2\frd_\ggo-\frac12\,(1-\epsilon_\ggo)=
\begin{cases}
0,&\mathfrak{g}=\mathfrak{o}_{2n+1}\,,\\
1,&\mathfrak{g}=\mathfrak{sp}_{2n}\,,\\
2,&\mathfrak{g}=\mathfrak{o}_{2n}\,.
\end{cases}
\end{equation}
We will note $\Theta(m)$ the Heaviside step function, defined on $\ZZ$ 
by the relations
\begin{equation}\label{Theta}
\Theta(m)=\begin{cases} 1,\quad m\geq 0\,,\\ 0,\quad m<0\,.\end{cases}
\end{equation}

\subsection{$RTT$ presentation}
The main subject of this paper  will be  a 
generic $\mathfrak{g}$-invariant integrable model 
which is described by the $N\times N$ monodromy matrix $T(u)$.
 The monodromy matrix satisfies the commutation relations 
 \begin{equation*}\label{RTT}
  R(u,v) \left( T(u)\otimes\mathbf{I} \right) \left( \mathbf{I}\otimes T(v) \right) =
  \left( \mathbf{I}\otimes T(v) \right) \left( T(u)\otimes\mathbf{I} \right) R(u,v) ,
\end{equation*}
with the $\mathfrak{g}$-invariant $R$-matrix \r{RABCD}. 
We assume that the 
monodromy matrix $T(u)$ is a Laurent series with respect 
to the formal parameter $u$ of the form 
\begin{equation}\label{T-Laut}
T(u)=\mathbf{I}+\sum_{m\geq 0}T[m]\,(u/c)^{-m-1}\,.
\end{equation}

Using the 
monodromy matrix entries $T_{i,j}(u)$ defined 
by 
\begin{equation}\label{T-ent}
T(u)=\sum_{ i,j\in \Ig} \Ee_{i,j} \cdot T_{i,j}(u)
\end{equation}
the commutation relation can be written as follows 
\begin{equation}\label{rtt}
\begin{split}
  \left[ T_{i,j}(u), T_{k,l}(v) \right] &=  
  \frac{c}{u-v}\left( T_{k,j}(v)T_{i,l}(u) - T_{k,j}(u) T_{i,l}(v) \right)\\
  &+\frac{c}{u-v+c\kappa_\ggo}\sum_{p\in \Ig}\epsilon_{p}\left(\delta_{k,i'}\,\epsilon_{i} \, T_{p,j}(u)T_{p'\!,l}(v)-
  \delta_{l,j'}\,\epsilon_{j}\,  T_{k,p'}(v)T_{i,p}(u)\right),
  \end{split}
\end{equation}
with  $i'$ and $\epsilon_{i}$ are defined by \eqref{prime} and \r{eps-i} 
respectively. 
Let us remind that the  second line in \r{rtt} is missing 
for the generic $\mathfrak{gl}_n$-invariant integrable model
since the parameters $\epsilon_i$ equal to 0
in this case.
It follows from the definitions \r{T-Laut} and \r{T-ent} 
that the monodromy entries $T_{i,j}(u)$ are Laurent series of the 
formal spectral parameter $u$ of the form 
\begin{equation}\label{T-eL}
T_{i,j}(u)=\delta_{i,j}+\sum_{m\geq 0}T_{i,j}[m]\,(u/c)^{-m-1}\,.
\end{equation}

We denote by $\Alm_\ggo$ the algebra of operators $T_{i,j}(u)$ satisfying 
the commutation relations \r{rtt}. We will loosely call $\Alm_\ggo$ the Yangian $Y(\ggo)$ of $\ggo$, although
the algebraically dependent 
generators of the Yangian $Y(\ggo)$ are the modes $T_{i,j}[m]$, $m\geq0$ 
of the 
generating series $T_{i,j}(u)$.  Among the modes there are the so-called
zero modes operators  $\TT_{i,j}=T_{i,j}[0]$. One can verify from \r{rtt} that
the algebra of zero modes operators $\TT_{i,j}$ is isomorphic 
to the symmetry algebra $\ggo$. Using \r{rtt} and \r{T-eL} we can write the commutation 
relations of the zero modes $\TT_{i,j}$ with the monodromy entries $T_{k,l}(v)$
 for all $\ggo$ 
\begin{equation}\label{rtt-zm-g}
\left[ \TT_{i,j}, T_{k,l}(v) \right] =  
  \Big(\delta_{i,l}\,T_{k,j}(v) -\epsilon_i\,\epsilon_j\,\delta_{j'\!,l}\,T_{k,i'}(v)\Big)
  - 
  \Big(\delta_{k,j}\,T_{i,l}(v) -\epsilon_i\,\epsilon_j\,\delta_{i'\!,k}\,T_{j'\!,l}(v)\Big).
\end{equation}

We fix a basis for the Cartan generators of $\ggo$ in the following way.
We first introduce the diagonal zero modes
\begin{equation}\label{k-zero}
\Sk_a=T_{a+1,a+1}[0]-\lambda_{a+1}[0]\,,\quad a\in J_\ggo,
\end{equation}
such that $\Sk_a\,\rvac=0$, where $\rvac$ is a vacuum vector of generalized integrable model defined below in \r{rvec}. 
It follows from 
the commutation relation \r{rtt-zm-g}   that the monodromy matrix 
entries $T_{i,j}(u)$ are eigenvectors with respect to adjoint action of 
the operators \r{k-zero} 
\begin{equation}\label{ad-dzm}
[\Sk_{a},T_{i,j}(z)]=\Big((\delta_{a+1,j}-\delta_{a+1,i})
-\epsilon^2_{\ggo}\,(\delta_{a+1,j'}-\delta_{a+1,i'})\Big)\,T_{i,j}(z)\,,
\end{equation}
where $\epsilon_\ggo$ is given in \r{eps-g}.

For $a\in\PSR_\ggo$, let $ \nus_a$ and $\fd_{a}$
 be the following parameters 
\begin{equation}\label{nusa}
\nus_a=1+(\frd_\ggo+(\epsilon_\ggo-1)/2)\,\delta_{a,0}=
\begin{cases}1,&\ggo=\gln,\,\ggb,\,\gsp\,,\\
1+\delta_{a,0},&\ggo=\ggd\,,
\end{cases}
\end{equation}
and 
\begin{equation*}\label{fd-def}
\fd_{a}=\epsilon^2_{\ggo}\,\frd_\ggo\, \sum_{\ell=0}^{\varphi_\ggo-1}\delta_{a,\ell}=\begin{cases}\,
0,\quad &\ggo=\gln,\,\ggb\,,\\
\delta_{a,0},\quad &\ggo=\gsp\,,\\
\delta_{a,0}+\delta_{a,1},\quad &\ggo=\ggd\,.
\end{cases}
\end{equation*}

Then we can define the {\it color operators} 
$\col_a$ as
\begin{equation}\label{col-ope-b}
\col_a=\frac{1}{1+\fd_{a}}\left(\sum_{b=a+\nus_a-1}^{n-1}
\Sk_b+(\varphi_\ggo-1)\,(-1)^a\,\fd_{a}\,\Sk_0 \right)\,,\quad a\in\PSR_\ggo\,.
\end{equation}
Note that, due to the values of $\epsilon_{\ggo}$, $\varphi_\ggo$ and $\xi_\ggo$,
the second term in  \eqref{col-ope-b}
is non-zero only for $\ggo=\ggd$. In that case 
this terms is equal to $(\delta_{a,0}-\delta_{a,1})\,\Sk_0$. Note also that for $\ggo=\gln$ all the parameters $\nus_a=1$, since 
 for this algebra $a=0$ does not occur.

Using \r{ad-dzm} one can calculate the adjoint action of the color operators 
$\col_a$, $a\in\PSR_\ggo$
on the monodromy entries $T_{i,j}(z)$: 
\begin{equation*}\label{ad-Tij}
[\col_a,T_{i,j}(z)]=\Big(q_a(j)-q_a(i)-\epsilon^2_\ggo\big(q_a(j')-q_a(i')\big)\Big)\,T_{i,j}(z)\,,
\end{equation*}
where 
\begin{equation*}\label{qbi}
q_a(i)=\frac{1}{1+\fd_{a}}\left( \sum_{b=a+\nus_a-1}^{n-1}
\delta_{b+1,i}+\delta_{\varphi_\ggo,2}\, 
(\delta_{a',i}-\delta_{a,i})\,\delta_{i,1}\right)\,.
\end{equation*}

Using properties of the $R$-matrix \r{RABCD} one can also verify that 
in the algebra $\Alm_{\mathfrak{g}}$ 
 with $\ggo\not=\mathfrak{gl}_n$ there 
are  symmetry relations 
\begin{equation}\label{sbcd}
T(u-c\,\kappa_{\mathfrak{g}})^{\rm t}\cdot T(u) = T(u)\cdot T(u-c\,\kappa_{\mathfrak{g}})^{\rm t} =
z(u)\ \mathbf{I}\,,
\end{equation}
 where $z(u)$ is a central element in $Y({\mathfrak{g}})$.

In what follows, and still for\footnote{Note that one can formally extend the relations \eqref{sbcd} by setting  $z(u)=0$ when $\ggo=\gln$, which makes them trivial.} $\ggo\neq \gln$,  we will fix this central element  by the relation 
$z(u)=1$ so that
the symmetry relations  simplify to 
\begin{equation}\label{sbcd1}
\Big(T(u)^{\rm t}\Big)^{-1}=T(u+c\kappa_{\mathfrak{g}})
\quad\mbox{or}\quad 
\Big(T(u)^{-1}\Big)^{\rm t}=T(u-c\kappa_{\mathfrak{g}})\,,
\end{equation}
which proves that both maps 
$T(u)\to(T(u)^{\rm t})^{-1}$ and 
$T(u)\to(T(u)^{-1})^{\rm t}$ are automorphisms 
of the algebra $\Alm_\ggo$.

In the framework of the universal algebraic Bethe ansatz 
a generic $\ggo$-invariant integrable model is characterized 
by the algebra $\Alm_\ggo$ and its representation space $\mathcal{H}$
generated by the vectors $\rvec$ such that 
\begin{equation}\label{rvec}
\LL_{i,j}(u)\rvec =0\,,\quad i>j\,,\quad \LL_{i,i}(u)\rvec=
\lambda_i(u)\rvec\,,\quad i\in\Ig\,.
\end{equation}
The vectors $\rvec$, when they exist, are called {\it vacuum vectors}. 
The functions $\lambda_i(u)$ are functional parameters of a generic 
integrable model. They are free for 
$\mathfrak{gl}_n$-invariant integrable models, while for $\mathfrak{o}_N$ and $\mathfrak{sp}_{2n}$-invariant
models they satisfy certain 
algebraic relations  which follow from the symmetry relation \r{sbcd1}.
Note also that these symmetry relations are a generalization of the relation for 
the classical orthogonal and symplectic matrices. 

The Bethe vectors $\BB(\bt)$ in a generic $\ggo$-invariant integrable models
depend on sets of  Bethe parameters $\bt^s$ 
of the cardinality $m_s\geq 0$ labeled by 
the simple roots $\rt_s$ of the algebra $\ggo$. 
The Bethe vectos are defined by a certain polynomials 
of noncommutative entries $T_{i,j}(u)$ for 
$i < j$ acting on the vacuum vector $\rvec$. Thus, they belong to the space
\begin{align*}\label{setV}
    \cV^+=\Big\{\mathcal{P}\big(T_{i_1, j_1}(u_1), \ldots, T_{i_k, j_k}(u_k)\big)\, \rvec,\quad 
     i_\ell < j_\ell,\quad
     i_\ell,j_\ell\in\Ig,\quad 
     1\le \ell \le k,\quad
    \ \mathcal{P} \in \mathcal{S}\Big\},
\end{align*}
where 
$(u_1, \ldots, u_k)$ is any collection of Bethe parameters
and $\mathcal{S}$ is the set of all polynomials whose coefficients are  rational functions in all the Bethe parameters $u_\ell$  and  
 the functionals parameters $\lambda_i(u_\ell)$, 
 $i\in\Ig$.

There are essentially two ways to define the Bethe vectors $\BB(\bar t)\in\cV^+$. The first method is to define the on-shell Bethe vectors as elements of $\cV^+$ such that when the Bethe equations (to be defined)  are obeyed they become  eigenvectors of the transfer matrix, see for example the nested Bethe ansatz approach \cite{KR1,KR2, DVK, Res85, Res91}. This is a loose definition for off-shell Bethe vectors, since any quantity which vanishes on the Bethe equations can be added without spoiling the eigenvector property. The second method gives a precise definition of off-shell Bethe vectors as some projection of a product of currents of the Yangian double $DY(\fg)$  \cite{HLPRS17, LP2, LPR-PM}. The definition is precise but has the drawback that one needs to work in the Drinfeld's new realization  (not the $RTT$ presentation) and moreover to be in the double of the Yangian 
$Y(\ggo)$. 

In the present paper, we introduce a precise definition of off-shell Bethe vectors within the $RTT$ presentation. The definition is inspired by some results obtained from the current presentation  \cite{LPR-PM}, but once the definition is fixed, all the properties for off-shell Bethe vectors are obtained using only the monodromy matrix $T(u)$, and one can forget about the Drinfeld's current presentation. 

\subsection{Rational functions}
Let $g(u,v)$, $h(u,v)$, $f(u,v)$ and $\gamma(u,v)$ be the rational functions 
\begin{equation*}\label{g-h-f-gl}
\begin{aligned}
h(u,v)&=\frac{u-v+c}{c},\qquad &g(u,v)&=\frac{c}{u-v}\,,\\
f(u,v)&=h(u,v)\,g(u,v),\qquad &\gamma(u,v)&=\frac{g(u,v)}{h(v,u)}\,.
\end{aligned}
\end{equation*}

In what follows we will use partitions of the sets of Bethe parameters 
$\bt^a\dashv\{\bt^a_{\so},\bt^a_{\st}\}$ 
into disjoint (possibly empty) subsets $\bt^a_{\so}\cap \bt^a_{\st}=\vn$
such that the cardinality $|\bt^a|=|\bt^a_{\so}|+|\bt^a_{\st}|$.
We will systematically use shorthand notations 
for the products of the functions depending on the Bethe parameters 
from the same set. For example, 
\begin{equation}\label{prod-ex}
g(\bt^a_{\so},\bt^b_{\st})=\prod_{t^a_i\in\bt^a_{\so}}\,
\prod_{t^b_j\in\bt^b_{\st}}\, g(t^a_i,t^b_j)\,.
\end{equation}
There is no singularities in \r{prod-ex} for $a=b$ since 
the subsets $\bt^a_{\so}$ and $\bt^a_{\st}$ are disjoint, which means that 
if $t^a_m\in\bt^a_{\so}$ then $t^a_m\not\in\bt^a_{\st}$.
We will use the convention that any product $\prod_{s=a}^b$ is equal to 1 if 
$a>b$.
We will also use the convention that when one of the sets is empty, then the product is 1, e.g.
\begin{equation*}
f(\vn,\bar t)=f(\bar t,\vn)=1.
\end{equation*}


 We consider sums over partitions subject to the condition that the cardinalities of certain subsets are fixed.
If, in such a sum, the cardinality of any subset is formally specified to be negative, we define the corresponding sum to be zero.

Let  $h_a(u,v)$, $f_a(u,v)$, and $\gamma_a(u,v)$ be the rational functions 
\begin{equation}\label{g-h-f-gen}
h_a(u,v)=\frac{u-v+c\,\hgo_{a,a}/2}{c},\qquad \quad\
f_a(u,v)=h_a(u,v)\,g(u,v)\,,
\end{equation}
\begin{equation*}\label{g-gen}
\gamma_a(u,v)=\frac{f_a(u,v)}{h_a(u,v)^{\zeta_a}\,h_a(v,u)^{\zeta_a}},\qquad
\zeta_a=\frac{(\hgo_{a,a}-1)(6-\hgo_{a,a})}{\hgo_{a,a}+2}\,,
\end{equation*}
where $\hgo_{a,a}$ is the norm of the simple root $\rt_a$.
Note that the function $\gamma_a(u,v)$ is  always equal to
$\frac{g(u,v)}{h_a(v,u)}$ except for the algebra 
$\ggo=\ggb$ and $a=0$.  In the latter case it is equal to 
$\gamma_0(u,v)=f_0(u,v)=\frac{u-v+c/2}{u-v}$.

For the simple roots $\rt_a$ such that $\hgo_{a,a}=2$ we will 
use simplified notations for the functions \r{g-h-f-gen}
\begin{equation*}\label{h-f-sim}
h_a(u,v)\Big|_{\hgo_{a,a}=2}=h(u,v),\qquad 
f_a(u,v)\Big|_{\hgo_{a,a}=2}=f(u,v)\,.
\end{equation*}

For $a\in\PSR_\ggo$ 
let $\varrho_a$ be the following parameter 
 \begin{equation*}\label{nusp}
\varrho_a=1+\delta_{a,0}\,
(1-\epsilon_\ggo)/2\,=
\begin{cases}1,&\ggo=\gln,\mathfrak{o}_{N}\,,\\
1+\delta_{a,0},&\ggo=\gsp\,.
\end{cases}
\end{equation*}

For any two simple roots $\rt_a$ and $\rt_b$ of the algebra
 $\ggo$  we define 
\begin{equation*}\label{gam-all}
\bgam^b_a(u,v)=
\begin{cases}
g(v,u)^{-1}+\varrho_a,&b>a,\quad 
\hgo_{a,b}\not=0\,,\\[2mm]
g(u,v)^{-1},&b<a,\quad \hgo_{a,b}\not=0\,,\\[2mm]
\displaystyle \gamma_a(u,v),&
a=b,\\[2mm]
1,&a\not=b,\quad\hgo_{a,b}=0.
\end{cases}
\end{equation*}

\section{The Bethe vectors for $\mathfrak{gl}_n$}
\label{sect4}

Let $\PSR_{\gln}$ be a set of  indices of the nodes for
 $\mathfrak{gl}_n$ Dynkin diagram \eqref{PSR-g}
\begin{equation*}\label{PSR-gln}
\PSR_{\gln}=\{1,2,\ldots,n-1\}\,.
\end{equation*}

For $a\in\PSR_{\mathfrak{gl}_n}$ and a set $\bt = \{\bt^1, \ldots, \bt^{n-1}\}$ we define the functions 
$\alpha_a(\bt^a)$ and $\alpha(\bt)$
\begin{equation}\label{alpha-gl}
\alpha_a(\bt^a)= \frac{\lambda_a(\bt^a)}{\lambda_{a+1}(\bt^a)},\qquad
\alpha(\bt)=\prod_{a\,\in\,\PSR_{\gln}}\,\alpha_a(\bt^a)\,.
\end{equation}

The generators of the nilpotent 
part of the algebra $\ggo$ are expressed in terms of the zero modes 
of the monodromy matrix as follows 
\begin{equation*}\label{zm-sr-gl}
\Tzm_a=\TT_{a+1,a}\,.
\end{equation*}

\begin{Def}\label{def:BV-gln}
The off-shell Bethe vectors in $\gln$ invariant models are the elements $\BB(\bar t)$ obeying the following properties
\begin{eqnarray}
\label{T-act-gl}
T_{1,n}(z)\cdot\BB(\bt) &=&\lambda_n(z)\,h(\bt^1,z)\,h(z,\bt^{n-1})\,\BB(\bar u)\,,\\
\label{zm-ac1}
\Tzm_a\cdot\BB(\bt)&=&\sum_{\rm part}
\Big(\alpha_a(\bt^a_{\so})\,\Oml_a(\bt_{\st},\bt_{\so})-
\Omr_a(\bt_{\so},\bt_{\st})\Big)\BB(\bt_{\st})\,,\quad a=1,2,...,n-1,
\end{eqnarray}
where we have introduced the extended sets  
\begin{equation}\label{exA}
 \bu = \{\bu^1, \ldots, \bu^{n-1}\}, \qquad \bar u^s=\{\bt^s,z\}\,.
\end{equation}
In the  equation \eqref{zm-ac1}, the sum runs over all partitions $\bar t^{a}\dashv\{\bar t^{a}_{\so},\bar t^{a}_{\st}\}$ with $|\bar t^{a}_{\so}| = 1$ and $\bar t^{b}_{\st} = \bar t^{b}$ for $b \ne a$,
and we used the functions
\begin{equation}\label{zm-ac2}
\begin{split}
\Omr_a(\bt_{\so},\bt_{\st})&=\prod_{b\,\in\,\PSR_{\gln}}
\bgam_a^b(\bt^a_{\so},\bt^b_{\st})=
\frac{g(\bt^a_{\so},\bt^a_{\st})}{h(\bt^a_{\st},\bt^a_{\so})}\,
\frac{h(\bt^{a+1}_{\st},\bt^a_{\so})}{g(\bt^a_{\so},\bt^{a-1}_{\st})}=
\Omr(\bt^a_{\so},\bt^a_{\st}|\bt^{a-1}_{\st},\bt^{a+1}_{\st})\,,\\
\Oml_a(\bt_{\st},\bt_{\so})&=\prod_{b\,\in\,\PSR_{\gln}}
\bgam_b^a(\bt^b_{\st},\bt^a_{\so})=
\frac{g(\bt^a_{\st},\bt^a_{\so})}{h(\bt^a_{\so},\bt^a_{\st})}\,
\frac{h(\bt^{a}_{\so},\bt^{a-1}_{\st})}{g(\bt^{a+1}_{\st},\bt^{a}_{\so})}=
\Oml(\bt^a_{\st},\bt^a_{\so}|\bt^{a-1}_{\st},\bt^{a+1}_{\st})\,.
\end{split}
\end{equation}
\end{Def}

\subsection{Action formulas and recurrence relations of $\gln$ type Bethe vectors}

 It was shown in the papers \cite{HLPRS20,LPR25} 
that it is sufficient to use the definition \ref{def:BV-gln}
to get the general action formulas and the general recurrence relation. 
Here we formulate these results without proofs, but recall some details of
the proof of the eigenvector property of on-shell Bethe vectors
using the action formulas of the diagonal monodromy entries,
also given in  \cite{HLPRS17}. 

In what follows we will often 
 use the notation $\{\bt^s\}_a^b$ for a collection of  sets 
\begin{equation}\label{eq: colect of set}    
    \{\bt^s\}_a^b = \{\bt^a,\bt^{a+1},\ldots, \bt^{b-1},\bt^b\}.
\end{equation}
We will understand the 
collection $\{\bt^s\}_a^b$ with $a>b$ as a collection of empty sets. 

We describe all the action formulas proved in \cite{HLPRS20} 
for a slightly different normalization of the off-shell Bethe vectors
 and the rectangular recurrence relations for the off-shell Bethe vectors 
 proved in \cite{LPR25} 
 for  a generic $\gln$-invariant  integrable model. 
 We first need to introduce 

\begin{itemize}
\item The operators $\Z_\ell^k$, $k \ge l$, which generate extensions of the sets of Bethe parameters:
\begin{equation}\label{Zop-gl}
\Z_\ell^k\cdot\BB(\bt)=\BB(\{\bt^s\}_1^{\ell-1},\{\bu^s\}_\ell^{k-1},\{\bt^s\}_k^{n-1})=
\BB(\{\bt^s\}_1^{\ell-1},\{\bt^s,z\}_\ell^{k-1},\{\bt^s\}_k^{n-1})\,.
\end{equation}
Note that, due to \eqref{eq: colect of set}, $\Z_k^k\cdot\BB(\bt) = \BB(\bt)$.

\item The rational functions $\psi_\ell(z;\bt)$, $\phi_k(z;\bt)$,
and $\mu^k_\ell(z;\bt)$ for $1\leq\ell\leq k\leq n$
\begin{equation}\label{p-p-m-gl}
\begin{split}
\psi_\ell(z;\bt)&=h(\bt^\ell,z)\, g(z,\bt^{\ell-1})\,,\\[2mm]
\phi_\ell(z;\bt)&=g(\bt^\ell,z)\, h(z,\bt^{\ell-1})\,,\\[2mm]
\mu^k_\ell(z;\bt)&= \psi_\ell(z;\bt)\, \phi_k(z;\bt)\,.
\end{split}
\end{equation}
\end{itemize}

\begin{prop}\label{ac-gln} \cite{HLPRS20} 
Provided that the equalities 
 \r{T-act-gl} and \r{zm-ac1} are valid, 
 the actions of 
the monodromy entries $T_{i,j}(z)$ for $1\leq i,j\leq n$
on the off-shell Bethe vectors
$\BB(\bt)$ in a generic $\gln$-invariant integrable model
are given by 
\begin{equation}\label{T-all-gln}
T_{i,j}(z)\cdot \BB(\bt)= \lambda_n(z)\,\mu^n_{1}(z;\bt)\,
\sum_{{\rm part}\ \bu^s}
\BB(\bu_{\st})\, \frac{\Om(\bu_{\so},\bu_{\st},\bu_{\sth})\,\alpha(\bu_{\sth})}
{\psi_1(z;\bu_{\so})\,\phi_n(z;\bu_{\sth})}\,,
\end{equation}
where  the sum goes over partitions 
of the extended sets $\bu^s=\{\bar t^s,z\}$
with cardinalities 
\begin{equation}\label{car-gln}
|\bu^s_{\so}|=\Theta(i-1-s),\qquad |\bu^s_{\sth}|=\Theta(s-j),\qquad 
|\bu^s_{\st}|=|\bu^s|-|\bu^s_{\so}|-|\bu^s_{\sth}|\,,
\end{equation} 
where $\Theta$ is the Heaviside function defined in \eqref{Theta}.
We have introduced
\begin{equation}\label{Om-full}
\Om(\bu_{\so},\bu_{\st},\bu_{\sth})=\Om(\bu_{\so},\bu_{\st})\,
\Om(\bu_{\so},\bu_{\sth})\,\Om(\bu_{\st},\bu_{\sth})=
\prod_{a,b\,\in\,\PSR_{\gln}}\bgam_a^b(\bu^a_{\so},\bu^b_{\st})\,
\bgam_a^b(\bu^a_{\so},\bu^b_{\sth})\,
\bgam_a^b(\bu^a_{\st},\bu^b_{\sth})\,,
\end{equation}
defined for  any disjoint partitions 
$\bu^a\dashv\{\bu^a_{\so},\bu^a_{\st},\bu^a_{\sth}\}$,  $a\in\PSR_{\gln}$.
The r.h.s. of \r{T-all-gln} depends on the indices $i$ and $j$ through 
cardinalities of the subsets $\bu^s_{\so}$, $\bu^s_{\st}$, 
and $\bu^s_{\sth}$, see \r{car-gln}. 
\end{prop}

\begin{prop}\label{rr-gln} \cite{LPR25}
Provided that the equalities 
 \r{T-act-gl} and \r{zm-ac1} are valid, 
 the off-shell Bethe vectors in $\gln$-invariant integrable models 
satisfy the rectangular recurrence relations
\begin{equation}\label{rr-gln1}
\Z^k_\ell\cdot\BB(\bt)=\frac{1}{\lambda_k(z)\,\mu^k_\ell(z;\bt)}\ 
\sum_{i=1}^\ell\sum_{j=k}^n\sum_{{\rm part}\ \bt}
\alpha(\bt_{\sth})\,
\Xi^{\ell,k}_{i,j}(z;\bt_{\so},\bt_{\st},\bt_{\sth})\,T_{i,j}(z)\cdot\BB(\bt_{\st})\,,
\quad 1\leq \ell< k\leq n
\end{equation}
where 
\begin{equation}\label{Xi-rr-gln}
\Xi^{\ell,k}_{i,j}(z;\bt_{\so},\bt_{\st},\bt_{\sth})=
\psi_\ell(z;\bt_{\so})\,
\phi_k(z;\bt_{\sth})\,\Om(\bt_{\so},\bt_{\st})\,
\Om(\bt_{\st},\bt_{\sth})
\end{equation}
and the sum in \r{rr-gln1} goes over partitions of the sets
$\bt^s\dashv\{\bt^s_{\so},\bt^s_{\st},\bt^s_{\sth})$ with cardinalities
\begin{equation}\label{oCgl}
|\bt^s_{\so}|=
\Theta(s-i)\,\Theta(\ell-s-1),\quad
|\bt^s_{\sth}|=
\Theta(j-s-1)\,\Theta(s-k),\quad |\bt^s_{\st}|=|\bt^s|-|\bt^s_{\so}|-|\bt^s_{\sth}|\,.
\end{equation}
The r.h.s. of \r{Xi-rr-gln} depends on the indices $i$ and $j$ 
through the cardinalities of the subsets 
$\bt^s_{\so}$, $\bt^s_{\st}$, and $\bt^s_{\sth}$ given by 
\r{oCgl}.  
\end{prop}

\subsubsection{Eigenvalue property of on-shell Bethe vectors}\label{eig-val-sect}

\begin{Def}\label{be-gl}
The off-shell Bethe vectors become on-shell Bethe vectors if the Bethe 
parameters satisfy the Bethe equations, which for $\gln$-invariant 
integrable model can be written in the form 
\begin{equation}\label{BEA}
\alpha_a(\bt^a_{\so})=\frac{\Omr_a(\bt_{\so},\bt_{\st})}{\Oml_a(\bt_{\st},\bt_{\so})}
=\prod_{b\,\in\,\PSR_{\gln}}
\frac{\bgam_a^b(\bt^a_{\so},\bt^b_{\st})}
{\bgam_b^a(\bt^b_{\st},\bt^a_{\so})}=
\frac{f(\bt^a_{\so},\bt^a_{\st})}{f(\bt^a_{\st},\bt^a_{\so})}\
\frac{f(\bt^{a+1},\bt^a_{\so})}{f(\bt^a_{\so},\bt^{a-1})}\,,\quad a=1,2,...,n-1.
\end{equation}
where we assume $\bt^0 =\bt^n = \varnothing$.  According to \r{zm-ac1} and \r{BEA} the
on-shell Bethe vectors are highest weight vectors
with respect to the nilpotent generators of the algebra $\gln$. 
\end{Def}

To verify the eigenvalue property of the on-shell Bethe vectors 
we need to rewrite the action of the diagonal monodromy entries
$T_{i,i}(z)$ given by the proposition~\ref{ac-gln} as a 
 sum over original set of the Bethe parameters $\bt^s$. 
We formulate this assertion in following proposition. 

\begin{prop}\label{diag-act-gln}
The action of the diagonal monodromy entries $T_{i,i}(z)$ on the 
off-shell Bethe vectors $\BB(\bt)$ given by the proposition~\ref{ac-gln} 
can be rewritten as
\begin{equation}\label{d-ac1-gl}
T_{i,i}(z)\cdot\BB(\bt)=\sum_{\ell=1}^{i}\sum_{k=i}^n
\sum_{{\rm part}\ \bt}\, \lambda_k(z)\,
\mu^k_\ell(z;\bt_{\st})\,\Z^k_\ell\cdot \BB(\bt_{\st})\,
\Ups_i^{\ell,k}(z;\bt_{\so},\bt_{\st},\bt_{\sth})\,,
\end{equation}
where the sum in \r{d-ac1-gl} goes over partitions 
$\bt^s\dashv\{\bt^s_{\so},\bt^s_{\so},\bt^s_{\so}\}$ such that the
cardinalities of the subsets are 
\begin{equation}\label{aco00}
|\bt^s_{\so}|=\Theta(s-\ell)\,\Theta(i-s-1),\quad
|\bt^s_{\sth}|=\Theta(k-1-s)\,\Theta(s-i),\quad
|\bt^s_{\st}|=|\bt^s|-|\bt^s_{\so}|-|\bt^s_{\sth}|\,.
\end{equation}

The functions $\Ups_i^{\ell,k}(z;\bt_{\so},\bt_{\st},\bt_{\sth})$ 
are defined by the equalities 
\begin{equation}\label{d-ac2-gl}
\Ups_i^{\ell,k}(z;\bt_{\so},\bt_{\st},\bt_{\sth})=
\frac{\lambda_n(z)}{\lambda_k(z)}\,
\frac{\mu^n_{1}(z;\bt)}{\mu^k_\ell(z;\bt_{\st})}\,
\frac{\Om(\bu_{\so},\bu_{\st},\bu_{\sth})\,\alpha(u_{\sth})}
{\psi_{1}(z;\bu_{\so})\,\phi_n(z;\bu_{\sth})}\,,
\end{equation}
where the subsets in the partitions of the extended sets 
$\bu^s\dashv\{\bu^s_{\so},\bu^s_{\st},\bu^s_{\sth}\}$ 
in \r{d-ac2-gl} are explicitly described by the equalities 
\begin{equation}\label{bea1}
\begin{aligned}
\bu^s_{\so}&=\{z\}, \quad
&\bu^s_{\st}&=\bt^s, \qquad\qquad
&\bu^s_{\sth}&=\vn,   \qquad
&&1\leq s\leq\ell-1\,,\\[2mm]
\bu^s_{\so}&=\{\crd{\bt^s_{\so}}{1}\},
&\bu^s_{\st}&=\{z,\crd{\bt^s_{\st}}{m_s-1}\},
&\bu^s_{\sth}&=\vn,
&&\ell\leq s\leq i-1\,,\\[2mm]
\bu^s_{\so}&=\vn,
&\bu^s_{\st}&=\{z,\crd{\bt^s_{\st}}{m_s-1}\},
&\bu^s_{\sth}&=\{\crd{\bt^s_{\sth}}{1}\},
&&i\leq s\leq k-1\,,\\[2mm]
\bu^s_{\so}&=\vn,
&\bu^s_{\st}&=\bt^s,
&\bu^s_{\sth}&=\{z\},
&&k\leq s\leq n-1\,.
\end{aligned}
\end{equation}

\end{prop} 

In \r{bea1} the integers under the subsets $\crd{\bt^s_{\st}}{m_s-1}$ 
indicates their cardinalities. If for some color $s$ the cardinality $m_s=1$ then 
$\crd{\bt^s_{\st}}{0}$ means that the subset $\bt^s_{\st}$
is empty, while if the cardinality $m_s=0$ then both 
subsets $\bt^s_{\st}$ and $\bt^s_{\sth}$ or 
$\bt^s_{\so}$ and $\bt^s_{\st}$ are empty.

\proof
The action of the diagonal monodromy entries $T_{i,i}(z)$ as given in formula 
\r{T-all-gln} can be written more explicitly in the form 
\begin{equation}\label{acuptr}
\begin{split}
T_{i,i}(z)\cdot\BB(\bt)&= \lambda_n(z)
\sum_{{\rm part}\ \bu^s} \mu_1^n(z;\bu_{\st})\,\BB(\bu_{\st})\,
h(\bu^j_{\sth},\bu^{j-1}_{\so})\,\times\\
&\quad\times  \prod_{s = 1}^{i-1} 
\Omr(\bu^s_{\so},\bu^s_{\st}|\bu^{s-1}_{\st},\bu^{s+1}_{\st})\ 
\prod_{s = i}^{n-1}
\alpha_s(\bu^s_{\sth})\ 
\Oml(\bu^s_{\st},\bu^s_{\sth}|\bu^{s-1}_{\st},\bu^{s+1}_{\st})\,.
\end{split}
\end{equation}
In the sum over partitions in \r{acuptr}, let us consider the two terms 
when $\bu^1_{\so}=\bt^1_{\so}$ and $\bu^1_{\so}=\{z\}$. 
In the term corresponding to  $\bu^1_{\so}=\bt^1_{\so}$, the function 
\begin{equation*}
\Omr(\bu^2_{\so},\bu^2_{\st}|\bu^{1}_{\st},\bu^{3}_{\st})=
\frac{g(\bu^2_{\so},\bu^2_{\st})}{h(\bu^2_{\st},\bu^2_{\so})}\ 
\frac{h(\bu^3_{\st},\bu^2_{\so})}{g(\bu^2_{\so},\{\bt^1_{\st},z\})}
\end{equation*}
will vanish when $\bu^2_{\so}=\{z\}$ since $g(z,z)^{-1}=0$ 
and in order to get non-vanishing contribution one has to 
require that  $\bu^2_{\so}=\bt^2_{\so}$. 
Continuing we obtain that if $\bu^1_{\so}=\bt^1_{\so}$ then 
$\bu^s_{\so}=\bt^s_{\so}$ for all $s>1$.

On the other hand when $\bu^1_{\so}=\{z\}$ the product of  functions 
\begin{equation*}
\frac{\mu^n_1(z;\bt)}{h(\bu^1_{\so},z)}\ 
\Omr(\bu^1_{\so},\bu^1_{\st}|\varnothing,\bu^{2}_{\st})=
\mu^n_1(z;\bt)\ \Omr(z,\bt^1|\varnothing,\bu^{2}_{\st})=
\frac{\mu^n_1(z;\bt)\ }{h(\bu^2_{\so},z)}\ \frac{g(z,\bt^1)}{h(\bt^1,z)}\ h(\bt^2,z)=
\frac{\mu^n_2(z;\bt)\ }{h(\bu^2_{\so},z)}
\end{equation*}
changes the normalization factor $\mu^n_1(z;\bt)\to \mu^n_2(z;\bt)$ with the
replacement $h(\bu^1_{\so},z)$ by $h(\bu^2_{\so},z)$
since $h(z,z)=1$. The subset $\bu^1_{\st}$ becomes the set $\bt^1$ 
and the sum over partitions of the set $\bu^1=\{\bt^1,z\}$ 
is absent. Now 
in the remaining sum over partitions we can consider two types of terms  
$\bu^2_{\so}=\bt^2_{\so}$ and $\bu^2_{\so}=\{z\}$ and repeat the arguments 
as above. 

Quite analogously we can show that if $\bu^{n-1}_{\sth}=\bt^{n-1}_{\sth}$
then all the terms in the sum over partitions of $\bu^s$ for $i\leq s\leq n-2$
vanish when $\bu^s_{\sth}=\{z\}$. But when $\bu^{n-1}_{\sth}=\{z\}$
the subset $\bu^{n-2}_{\sth}$ can be either $\bt^{n-2}_{\sth}$ or $\{z\}$. 

Gathering the above results, we can introduce two integers $k,\ell$ with 
$1\leq \ell\leq i\leq k\leq n$  such that 
$\bu^s_{\so}=\{z\}$, $\bu^s_{\st}=\bt^s$ for $1\leq s< \ell$ and 
$\bu^s_{\so}=\bt^s_{\so}$ for $\ell\leq s< i$, while
$\bu^s_{\sth}=\{z\}$, $\bu^s_{\st}=\bt^s$ for $k\leq s\leq n-1$ and 
$\bu^s_{\sth}=\bt^s_{\sth}$ for $i\leq s< k$. 
Then one can calculate 
\begin{equation*}\label{aco7}
\begin{split}
\lambda_n(z)\,\mu_1^n(z;\bu_{\st})\ 
&\prod_{s=1}^{\ell-1}\Omr(\bu^{s}_{\so},
\bu^{s}_{\st}|\bu^{s-1}_{\st},\bu^{s+1}_{\st})
\prod_{s=k}^{n-1}\alpha_s(\bu^s_{\sth})
\Oml(\bu^{s}_{\st},\bu^{s}_{\sth}|\bu^{s-1}_{\st},\bu^{s+1}_{\st})=\\
&=
\frac{\lambda_k(z)\,g(z,\bt^{\ell-1})\,h(\bt^\ell,z)\,
h(z,\bt^{k-1})\,g(\bt^k,z)}{h(\bt^\ell_{\so},z)\,h(z,\bt^{k-1}_{\sth})}=
\lambda_k(z)\, \mu^k_\ell(z;\bt_{\st})\,,
\end{split}
\end{equation*}
\begin{equation}\label{aco8}
\prod_{s=\ell}^{i-1}
\Omr(\bu^{s}_{\so},\bu^{s}_{\st}|\bu^{s-1}_{\st},\bu^{s+1}_{\st})=
g(\bt_{\so}^\ell,z)\,
\prod_{s=\ell}^{i-1}
\Omr(\bt^{s}_{\so},\bt^{s}_{\st}|\bt^{s-1}_{\st},\bt^{s+1}_{\st})\,,
\end{equation}
and 
\begin{equation}\label{aco9}
\prod_{s=i}^{k-1}
\Oml(\bu^{s}_{\st},\bu^{s}_{\sth}|\bu^{s-1}_{\st},\bu^{s+1}_{\st})=
g(z,\bt_{\sth}^{k-1})\,
\prod_{s=i}^{k-1}
\Oml(\bt^{s}_{\st},\bt^{s}_{\sth}|\bt^{s-1}_{\st},\bt^{s+1}_{\st})\,.
\end{equation}
Note that there is no contradictions in the formulas \r{aco8} or 
\r{aco9} when $i=\ell$ or $i=k$ since  according to the cardinalities 
\r{aco00} for these 
values of  $i$ 
one gets $\bt^\ell_{\so}=\vn$ or $\bt^{k-1}_{\sth}=\vn$.

The summation over partitions of the sets $\{\bt^s,z\}$ 
in \r{bea1} is reduced to the summation over 
$\ell$, $k$ and the sum over partitions of the sets 
$\bt^s\dashv\{\bt^s_{\so},\bt^s_{\st},\bt^s_{\sth}\}$ 
with cardinalities given by \r{aco00}. 

The equality \r{acuptr} becomes the equality \r{d-ac1-gl} with 
the function $\Ups_i^{\ell,k}(z;\bt_{\so},\bt_{\st},\bt_{\sth})$ equal to 
\begin{equation}\label{aco6}
\begin{split}
\Ups_i^{\ell,k}(z;\bt_{\so},\bt_{\st},\bt_{\sth})&=  
h(\bt^i_{\sth},\bt^{i-1}_{\so})\
g(\bt_{\so}^\ell,z)\ 
g(z,\bt_{\sth}^{k-1})\times\\
&\quad\times  \prod_{s = \ell}^{i-1} 
\Omr(\bt^s_{\so},\bt^s_{\st}|\bt^{s-1}_{\st},\bt^{s+1}_{\st})
 \prod_{s = i}^{k-1}
\alpha_s(\bt^s_{\sth})\ \Oml(\bt^s_{\st},\bt^s_{\sth}|\bt^{s-1}_{\st},\bt^{s+1}_{\st})\,.
\end{split}
\end{equation}
The fact that the r.h.s. of \r{aco6} coincides with the r.h.s. of \r{d-ac2-gl} 
is verified by a direct substitution of the subsets \r{bea1} in 
the r.h.s. of \r{d-ac2-gl}.\qed

\begin{cor}
The off-shell Bethe vectors have definite colors:
\begin{equation}
\col_{a}\, \BB(\bar t) = |\bar t^a|\, \BB(\bar t)\,, \quad a\in J_{\mathfrak{gl}_n}\,.
\end{equation}
\end{cor}
\proof
Considering the action formula \eqref{d-ac1-gl} in the limit $z\to\infty$, the coefficient of $\frac1z$ provides the result.\qed

Using the results of proposition~\ref{diag-act-gln} we can now calculate the 
action of the transfer matrix 
\begin{equation}\label{trans-gln}
\mathcal{T}_{\gln}(z)=\sum_{i=1}^n\,T_{i,i}(z)
\end{equation}
on the off-shell Bethe vectors $\BB(\bt)$
\begin{equation}\label{tr-ac-gln}
\begin{split}
\mathcal{T}_{\gln}(z)\cdot \BB(\bar t)&=
\sum_{i=1}^n
\sum_{\ell=1}^{i}\sum_{k=i}^n 
\sum_{{\rm part}\ \bt} \lambda_k(z)\,
\mu_{\ell}^k(z;\bt_{\st})\,\Z_\ell^k\cdot\BB(\bt_{\st})
 \Ups_{i}^{\ell,k}(z;\bt_{\so},\bt_{\st},\bt_{\sth})=\\
 &=\BB(\bt)\,\sum_{i=1}^n\lambda_i(z)\,
\mu_i^i(z;\bt)+\\
&+\sum_{\ell=1}^{n-1}\sum_{k=\ell+1}^n 
\sum_{{\rm part}\ \bt}\lambda_k(z)\, \mu_{\ell}^k(z;\bt_{\st})\,\Z_\ell^k\cdot\BB(\bt_{\st})
\sum_{i=\ell}^k \Ups_{i}^{\ell,k}(z;\bt_{\so},\bt_{\st},\bt_{\sth})\,.
\end{split}
\end{equation}
We can exchange 
the order of summation in \r{tr-ac-gln} because the subsets 
$\bt^s_{\st}$ do not depend on $i$ when $\ell<k$ as it can be seen explicitly  in \r{bea1}.

Due to the explicit form of the function 
$\Ups_{i}^{\ell,k}(z;\bt_{\so},\bt_{\st},\bt_{\sth})$ given 
by \r{aco6} the action of the transfer matrix 
\r{tr-ac-gln}  explicitly describes the structure of all wanted and unwanted 
terms in this action. If the Bethe parameters are generic the sum 
$\sum_{i=\ell}^k \Ups_{i}^{\ell,k}(z;\bt_{\so},\bt_{\st},\bt_{\sth})$ in \r{tr-ac-gln}
does not vanish. But if Bethe parameters satisfies the Bethe equations 
\r{BEA}, the function $\Ups_{i}^{\ell,k}(z;\bt_{\so},\bt_{\st},\bt_{\sth})$
\r{aco6} can be rewritten, for different values of $i$ and 
according to the cardinalities  \r{aco00},  as 
\begin{itemize}

\item $i=\ell$, we have $|\bt^s_\so|=0$, $|\bt^s_\sth|=1$ for $\ell\leq s\leq k-1$, leading to
\begin{equation*}\label{aco66a}
\Ups_\ell^{\ell,k}(z;\vn,\bt_{\st},\bt_{\sth})=  
g(z,\bt_{\sth}^{k-1})\,  \prod_{s = \ell}^{k-1} 
\Omr(\bt^s_{\sth},\bt^s_{\st}|\bt^{s-1}_{\st},\bt^{s+1}_{\st})\,.
\end{equation*}

\item $i=k$, we have $|\bt^s_\so|=1$, $|\bt^s_\sth|=0$ for $\ell\leq s\leq k-1$, leading to
\begin{equation*}\label{aco66b}
\Ups_k^{\ell,k}(z;\bt_{\so},\bt_{\st},\vn)=  
g(\bt_{\so}^\ell,z)\,  \prod_{s = \ell}^{k-1} 
\Omr(\bt^s_{\so},\bt^s_{\st}|\bt^{s-1}_{\st},\bt^{s+1}_{\st})\,.
\end{equation*}

\item $\ell<i<k$, we have $|\bt^s_\so|=1$, $|\bt^s_\sth|=0$ for $\ell\leq s\leq i-1$ and 
$|\bt^s_\so|=0$, $|\bt^s_\sth|=1$, $i\leq s\leq k-1$
\begin{equation}\label{aco66c}
\Ups_i^{\ell,k}(z;\bt_{\so},\bt_{\st},\bt_{\sth})=  
g(\bt^i_{\sth},\bt^{i-1}_{\so})^{-1}\,
g(\bt_{\so}^\ell,z)\, 
g(z,\bt_{\sth}^{k-1})\,  \prod_{s = \ell}^{k-1} 
\Omr(\bt^s_{\so,\sth},\bt^s_{\st}|\bt^{s-1}_{\st},\bt^{s+1}_{\st})\,.
\end{equation}

In \r{aco66c} the subset $\bt^s_{\so,\sth}$ is equal to $\bt^s_\so$ for 
$\ell\leq s\leq i-1$ and to the subset $\bt^s_{\sth}$ for 
$i\leq s\leq k-1$.
\end{itemize}

Since the cardinalities of all non-empty subsets $\bt^s_\so$ and $\bt^s_\sth$ 
are equal to 1, in the sum over partitions in \r{tr-ac-gln}, we can rename $\bt^s_\sth\to\bt^s_\so$ for the non-empty  subsets $\bt^s_\sth$.
Then the sum $\sum_{i=\ell}^k \Ups_{i}^{\ell,k}(z;\bt_{\so},\bt_{\st},\bt_{\sth})$
becomes proportional to the sum 
\begin{equation*}\label{vanish-id}
g(\bt^\ell_\so,z)+g(z,\bt^{k-1}_\so)+ g(\bt^\ell_\so,z)\,g(z,\bt^{k-1}_\so)
\sum_{i=\ell+1}^{k-1}g(\bt^i_\so,\bt^{i-1}_\so)
\end{equation*}
which identically vanishes.

This proves that if the Bethe parameters 
satisfy the Bethe equations then all unwanted terms in the action of 
transfer matrix on the off-shell Bethe vectors disappear. 
According to definition~\ref{be-gl} off-shell Bethe vector becomes on-shell 
 and according to \r{tr-ac-gln} it becomes an eigenvector 
 of the transfer matrix \r{trans-gln} with the eigenvalue 
 \begin{equation}\label{BEA9}
\tau_{\gln}(z;\bt)=\sum_{i=1}^n\lambda_i(z)\,f(z,\bt^{i-1})\,f(\bt^i,z)\,.
\end{equation}

\section{The Bethe vectors for $\ggo=\mathfrak{o}_N$, 
$\mathfrak{sp}_{2n}$ }
\label{sect5}

In the rest of the paper we will demonstrate that the same program 
can be realized for the Bethe vectors in $\mathfrak{o}_N$ and 
$\mathfrak{sp}_{2n}$-invariant integrable models.

Let $\PSR_\ggo$ be a set of indices  of the nodes of
Dynkin diagram for $\ggo$ \eqref{PSR-g}
\begin{equation*}\label{PSR}
\PSR_\ggo=\{0,1,2,\ldots,n-1\}\,.
\end{equation*}

We  introduce the sign factor
\begin{equation}\label{sigma}
\sigma_i=2\Theta(i-1)-1=(-1)^{\delta_{i<1}}=\begin{cases}
1,& i>0\,,\\
-1,& i\leq 0\,,
\end{cases}
\end{equation}

For $a\in\PSR_\ggo$ we define the functions 
$\alpha_a(z)$ 
\begin{equation}\label{alpha-all}
\alpha_a(z)= \frac{\lambda_a(z)}{\lambda_{a+\nu_a}(z)}\,,
\end{equation}
where $\nus_a$ is given by \r{nusa}.
The simple root generators of the simple Lie algebra $\ggo$ 
are expressed in terms of the zero modes 
of the monodromy matrix as follows 
\begin{equation*}\label{zm-sr}
\Tzm_a=\TT_{a+\nu_a,a}\,.
\end{equation*}

For $\ggo\not=\mathfrak{gl}_n$ and 
$s\in\PSR_\ggo$,
let $z_s$ be the shifted parameter 
\begin{equation}\label{z-sh1}
z_s=z-c(s+\theta_\ggo),\qquad \bvphi_\ggo\leq s\leq n-1\,,
\end{equation} 
where $\theta_\ggo$ and $\bvphi_\ggo$ are defined in \eqref{t-vp} and \eqref{vphi} respectively. We also introduce the sets $\bz_s$ of  cardinality $1+\delta_{s\geq\bvphi_\ggo}$ and
defined by the equalities 
\begin{equation}\label{z-sh2}
\bz_s=\begin{cases}
\{z\}, & 0\leq s<\bvphi_\ggo\,,\\
\{z,z_s\},&\bvphi_\ggo\leq s\leq n-1\,.
\end{cases}
\end{equation}

\begin{Def}\label{def:BV-osp}
The off-shell Bethe vectors in $\ggo$ invariant models are the elements $\BB(\bar t)\in\cV^+$ obeying the following properties
\begin{eqnarray}
T_{n',n}(z)\cdot \BB(\bt)&=&\lambda_n(z)\,\mu^n_{n'}(z;\bt)\ \BB(\bw)\,,
\label{T-act-g}\\
\Tzm_a\cdot\BB(\bt)&=&\sum_{\rm part}
\Big(\alpha_a(\bt^a_{\so})\,\Oml_a(\bt_{\st},\bt_{\so})-
\Omr_a(\bt_{\so},\bt_{\st})\Big)\BB(\bt_{\st})\,,\quad a=0,1,2,...,n-1,\quad
\label{bvde1}
\end{eqnarray}
where we have introduced the extended sets  
\begin{equation}\label{exBCD}
 \bw = \{\bw^1, \ldots, \bw^{n-1}\}, \qquad \bar w^s=\{\bt^s,\bz_s\}
\end{equation}
and we used the functions
\begin{equation}\label{bvde2}
\Omr_a(\bt_{\so},\bt_{\st})=\prod_{b\,\in\,\PSR_{\ggo}}
\bgam_a^b(\bt^a_{\so},\bt^b_{\st}),\qquad
\Oml_a(\bt_{\st},\bt_{\so})=\prod_{b\,\in\,\PSR_{\ggo}}
\bgam_b^a(\bt^b_{\st},\bt^a_{\so})\,.
\end{equation}
\begin{equation*}\label{ovs-all}
\mu^n_{n'}(z;\bt)= \begin{cases}
\displaystyle  -\,\kappa_{\mathfrak{o}_{2n+1}}\ 
\frac{g(z-c/2,\bt^0)}{h(z,\bt^0)}\,
\frac{h(z,\bt^{n-1})}{g(z_n,\bt^{n-1})},&\ggo=\mathfrak{o}_{2n+1}\,,\\[4mm]
\displaystyle  -\,\kappa_{\mathfrak{sp}_{2n}}\ 
(-1)^{|\bt^0|}\,h(\bt^1,z)\,
\frac{h(z,\bt^{n-1})}{g(z_n,\bt^{n-1})},&\ggo=\mathfrak{sp}_{2n}\,,\\[4mm]
\displaystyle  -\,\kappa_{\mathfrak{o}_{2n}}\ 
(-1)^{|\bt^0|+|\bt^1|}\,
\frac{h(z,\bt^{n-1})}{g(z_n,\bt^{n-1})},&\ggo=\mathfrak{o}_{2n}\,.
\end{cases}
\end{equation*}
In the equation \eqref{bvde1}, the sum runs over all partitions $\bar t^{a}\dashv\{\bar t^{a}_{\so},\bar t^{a}_{\st}\}$ with $|\bar t^{a}_{\so}| = 1$ and $\bar t^{b}_{\st} = \bar t^{b}$ for $b \ne a$.
\end{Def}

\subsection{The monodromy entries action on off-shell Bethe vectors}

For any two sets $\bx=\{x_1,x_2\}$ and $\by=\{y_1,y_2\}$ 
of  cardinalities 2 we introduce the polynomial
\begin{equation}\label{phi-fun}
\phf(\bx|\by)=h(x_1,x_2)+\frac{h(x_1,y_1)}{g(x_2,y_2)}+
\frac{h(x_1,y_2)}{g(x_2,y_1)}\,.
\end{equation} 
Note that this  polynomial is symmetric 
with respect to permutations of the variables within the set 
$\bx$ and to permutations within $\by$. 
In what follows we will also need analogs of the polynomial
\r{phi-fun} when the cardinalities of one or both sets $\bx$ and 
$\by$ are less than 2. We set 
\begin{equation*}\label{p-f-pr}
\begin{split}
\phf(x_1,\varnothing|y_1,y_2)&=
1+\frac{1}{g(x_1,y_1)}+\frac{1}{g(x_1,y_2)}\,,\\
\phf(x_1,\vn|y_1,\vn)&=\phf(\vn,x_2|\vn,y_2)=1,\qquad \phf(\vn,\vn|\vn,\vn)=1\,.
\end{split}
\end{equation*}
Note that the polynomial $\phf(\bx|\by)$ 
is always equal to 1 when both sets $\bx$ and $\by$ have cardinality 
less than 1.   
The same is valid for the function $h(\by,\by)$ when the cardinality 
$|\by|\leq 1$. 

The polynomial $\phf(\bx|\by)$ is relevant only for the algebra 
$\ggo=\mathfrak{sp}_{2n}$. For any integer $\ell$ such that $n'\leq\ell\leq n$ 
and for any partition of the extended sets 
$\bw^a\dashv\{\bw^a_{\so},\bw^a_{\st},\bw^a_{\sth}\}$ 
we introduce  the functions
\begin{equation}\label{phf-sp}
\ophf_\ell(\bw_{\so})=\prod_{s=1}^{\ell-1}
\frac{\phf(\bw^{s+1}_{\so}|\bw^{s}_{\so})}{h(\bw^{s}_{\so},\bw^{s}_{\so})},
\qquad 
\ophf_\ell(\bw_{\sth})=\prod_{s=1}^{\ell-1}
\frac{\phf(\bw^{s+1}_{\sth}|\bw^{s}_{\sth})}{h(\bw^{s}_{\sth},\bw^{s}_{\sth})}\,.
\end{equation}
When $\ell=n$ and correspondingly $s=n-1$
 in the products over $s$ in the definition \r{phf-sp},
the subsets $\bw^n_{\so}$ and $\bw^n_{\sth}$ are fixed to the 
 sets $\{z_n\}$ and $\{z\}$ respectively.

For all algebras $\ggo$ 
we define the functions $\coA_i(\bw_{\so})$ and $\roA_i(\bw_{\sth})$
such that 
\begin{itemize}

\item for $\ggo=\mathfrak{o}_{N}$:
\begin{equation*}\label{oAoN}
\coA_i(\bw_{\so})=\sigma_{i-\frd_\ggo},\qquad 
\roA_j(\bw_{\sth})=\sigma_{j'-\frd_\ggo}\,,
\end{equation*}

\item for $\ggo=\mathfrak{sp}_{2n}$:
\begin{equation*}\label{oAsp}
\coA_i(\bw_{\so})=
\sigma_{i'}\,\ophf_{i}(\bw_{\so}),\qquad 
\roA_j(\bw_{\sth})=\ophf_{j'}(\bw_{\sth})\,.
\end{equation*}

\end{itemize}

Let $\phi_n(z;\bw_{\sth})$ be the function which is equal to $h(z,\bw^{n-1}_{\sth})$
for all $\ggo$.  Let $\psi_{n'}(z;\bw_{\so})$ be the function which is equal to  
$g(z_n,\bw^{n-1}_{\so})^{-1}$. 
These functions are particular examples of the 
functions $\psi_\ell(z;\bt)$ and $\phi_\ell(z;\bt)$
which are defined in appendix~\ref{AppB} for $n'\leq\ell\leq n$.

\begin{prop}\label{ac-main}
Provided that equalities 
\r{bvde1} and \r{T-act-g} are valid, 
 the actions of 
the monodromy entries $T_{i,j}(z)$ for $n'\leq i,j\leq n$
on the off-shell Bethe vectors
$\BB(\bt)$ read
\begin{equation}\label{T-act-all}
T_{i,j}(z)\cdot \BB(\bt)=\lambda_n(z)\,\mu^n_{n'}(z;\bt)\,
\sum_{{\rm part}\ \bw^s}
\BB(\bw_{\st})\, \frac{\coA_i(\bw_{\so})}{\psi_{n'}(z;\bw_{\so})}\, 
\frac{\roA_j(\bw_{\sth})}{\phi_n(z;\bw_{\sth})}\, 
\Om(\bw_{\so},\bw_{\st},\bw_{\sth})\,  
\alpha(\bw_{\sth})\,.
\end{equation}
In \r{T-act-all} the sum goes over partitions given 
by the equalities 
\begin{equation}\label{car-g}
\begin{split}
|\bw^s_{\so}|&=\begin{cases}
\Theta(i-\bvphi_\ggo)+\delta_{i',s}\,
\delta_{\bvphi_\ggo,2},&0\leq s<\bvphi_\ggo\,,\\
\Theta(i+s-\frd_{\ggo})
+\Theta(i-s-1),&\bvphi_\ggo\leq s\leq n-1\,,\\
\end{cases}
\\
|\bw^s_{\sth}|&=\begin{cases}
\Theta(j'-\bvphi_\ggo)+\delta_{j,s}\,
\delta_{\bvphi_\ggo,2},&0\leq s<\bvphi_\ggo\,,\\
\Theta(j'+s-\frd_{\ggo})+\Theta(j'-s-1)
,&\bvphi_\ggo\leq s\leq n-1\,.
\end{cases}
\end{split}
\end{equation} 
\end{prop}

\proof 
The proof is based on an induction 
which uses the action \r{bvde1} of zero modes operators $\Tzm_a$, $a\in\PSR_\ggo$
on the off-shell Bethe vectors 
and the commutation relation of these zero mode operators
with monodromy entries $T_{i,j}(z)$ \r{rtt-zm-g}
\begin{equation}\label{zT-g}
[\Tzm_a,T_{i,j}(z)]=
(\delta_{a,j-1}-\epsilon_a\,\delta_{a,j'})\,T_{i,j-\nus_a}(z)-
(\delta_{a,i}-\epsilon_a\,\delta_{a,i'-1})\,T_{i+\nus_a,j}(z)\,.
\end{equation}

The starting point of the induction is the action  \r{T-act-g}
of the highest monodromy entry. Then, the application of the appropriate zero mode 
operator yields the action formulas for the values of the indices $n'<i\leq n$ 
and $n'\leq j<n$. 
These calculations are rather lengthy but straightforward and we leave
this exercise to the interested readers.
\qed

\begin{rem}
One can check that if the set $\bt^0$ in \r{T-act-all} is empty then the 
sum over partitions in the action of the monodromy entries $T_{1,n}(z)$
reduces to a single term and coincide with the action \r{T-act-gl} of 
the highest monodromy in $\gln$-invariant integrable model. 
Analogously, the action of zero modes operators $\Tzm_a$ for $a=1,\ldots,n-1$
\r{bvde1} becomes the action \r{zm-ac1} on the off-shell Bethe vector
$\BB(\vn,\{\bt^s\}_1^{n-1})$. This proves that the action formulas \r{T-act-all}
for $1\leq i,j\leq n$ for $\ggo\not=\gln$ becomes the action formulas 
\r{T-all-gln} of $\gln$ type. 
The same property also applies to the rectangular relations described below in section \ref{sec:rec-rec}.
\end{rem}

Note that when $i=n'$ and $j=n$ then according to the cardinalities 
\r{part-all} all subsets $\bw^s_{\so}=\bw^s_{\sth}=\vn$ are empty and 
$\coA_{n'}(\vn)\,\roA_n(\vn)=1$ for all $\ggo$. 
Moreover one can see that $\Om(\vn,\bw,\vn)=1$ and  by default  
$\alpha(\vn)=1$. 
It means that the sum over partitions 
in \r{T-act-all} disappear and this equality coincides with the action 
of the highest monodromy entry \r{T-act-g}.  

\begin{rem}
One can see from \r{car-g} that the 
cardinalities of the subsets $\bw^s_{\sth}$ can be obtained 
from the cardinalities $\bw^s_{\so}$ by the formal replacement 
$i\to j'$. 
Note also that the first lines in \r{car-g} are absent for $\ggo=\mathfrak{o}_{2n+1}$
because in these cases $\bvphi_\ggo$ is equal to $0$.
Notice that the cardinalities of the subsets 
$\bw^s_{\so}$ depend solely on the index $i$, while
the cardinalities of $\bw^s_{\sth}$ depend solely on the index $j$.
\end{rem}

For further use, we write explicitly the cardinalities of the partitions \r{car-g} 
for each $\ggo$.
\begin{itemize}

\begin{subequations}\label{part-all}

\item For $\ggo=\mathfrak{o}_{2n+1}$, $\bvphi_{\mathfrak{o}_{2n+1}}=0$,
$\frd_{\mathfrak{o}_{2n+1}}=0$:
\begin{equation}\label{partB}
\begin{aligned}
|\bw^s_{\so}|&=\Theta(i+s)+\Theta(i-s-1), &&s=0,\ldots,n-1\,,\\
|\bw^s_{\sth}|&=\Theta(-j+s)+\Theta(-j-s-1),\quad &&s=0,\ldots,n-1\,.
\end{aligned}
\end{equation}

\item For $\ggo=\mathfrak{sp}_{2n}$, $\bvphi_{\mathfrak{sp}_{2n}}=1$,
$\frd_{\mathfrak{sp}_{2n}}=1$:
\begin{equation}\label{partC}
\begin{split}
|\bw^s_{\so}|&=\begin{cases}
\Theta(i-1),&s=0\,,\\
\Theta(i+s-1)+\Theta(i-s-1),&s=1,\ldots,n-1\,,
\end{cases}
\\
|\bw^s_{\sth}|&=
\begin{cases}
\Theta(-j),&s=0\,,\\
\Theta(-j+s)+\Theta(-j-s),\quad\ \ &s=1,\ldots,n-1\,.
\end{cases}
\end{split}
\end{equation}

\item For $\ggo=\mathfrak{o}_{2n}$, $\bvphi_{\mathfrak{o}_{2n}}=2$,
$\frd_{\mathfrak{o}_{2n}}=1$:
\begin{equation}\label{partD}
\begin{split}
|\bw^s_{\so}|&=\begin{cases}
\Theta(i-2)+\delta_{i',s},&s=0,1\,,\\
\Theta(i+s-1)+\Theta(i-s-1),&s=2,\ldots,n-1\,,\\
\end{cases}
\\
|\bw^s_{\sth}|&=\begin{cases}
\Theta(-j-1)+\delta_{j,s},         &s=0,1\,,\\
\Theta(-j+s)+\Theta(-j-s),\quad\ \ &s=2,\ldots,n-1\,.\\
\end{cases}
\end{split}
\end{equation}
\end{subequations}

\end{itemize}

Using the definition \r{nusa} of the integer parameter $\nus_s$, 
the commutation relations \r{zT-g} can be also explicitly written
for different 
$\ggo$ 
\begin{itemize}
\begin{subequations}\label{zT-all}

\item for $\ggo=\mathfrak{o}_{2n+1}$, $0\leq a\leq n-1$:
\begin{equation}\label{zTo2n1}
[\Tzm_a,T_{i,j}(z)]=
(\delta_{a,j-1}-\delta_{a,-j})\,T_{i,j-1}(z)-
(\delta_{a,i}-\delta_{a,-i-1})\,T_{i+1,j}(z)\,,
\end{equation}

\item for $\ggo=\mathfrak{sp}_{2n}$, $0\leq a\leq n-1$:
\begin{equation}\label{zTsp}
[\Tzm_a,T_{i,j}(z)]=
(\delta_{a,j-1}-(-1)^{\delta_{a,0}}\,\delta_{a,-j+1})\,T_{i,j-1}(z)-
(\delta_{a,i}-(-1)^{\delta_{a,0}}\,\delta_{a,-i})\,T_{i+1,j}(z)\,,
\end{equation}

\item for $\ggo=\mathfrak{o}_{2n}$, $0\leq a\leq n-1$:
\begin{equation}\label{zTo2n}
[\Tzm_a,T_{i,j}(z)]=
(\delta_{a,j-\nus_a}-\delta_{a,-j+1})\,T_{i,j-\nus_a}(z)-
(\delta_{a,i-\nus_a}-\delta_{a,-i+1-\nus_a})\,T_{i+\nus_a,j}(z)\,.
\end{equation}

\end{subequations}
\end{itemize}

\subsection{Eigenvalue property of on-shell Bethe vectors}

To describe the eigenvalue property of on-shell Bethe vectors 
one has to consider the action of the diagonal entries $T_{i,i}(z)$ 
given by the proposition~\ref{ac-main} and calculate 
the action of the universal transfer matrix 
\begin{equation}\label{trans}
\mathcal{T}_\ggo(z)=\sum_{i=n'}^n\,T_{i,i}(z)\,.
\end{equation}
The calculation is done for on-shell Bethe vectors $\BB(\bt)$ which means that the sets of the Bethe 
parameters $\bt^s$ satisfy the Bethe equations 
\begin{equation}\label{BE-all}
\alpha_a(\bt^a_{\so})=\prod_{b\,\in\,\PSR_{\ggo}}
\frac{\bgam_a^b(\bt^a_{\so},\bt^b_{\st})}
{\bgam_b^a(\bt^b_{\st},\bt^a_{\so})}\,.
\end{equation}

To perform this calculation we will rewrite the action of diagonal 
monodromy entries from the sum over partitions of the extended sets 
$\bw^s$ to the sum over partitions of the original sets $\bt^s$ 
of the Bethe parameters. The action of the diagonal entries 
will produce the so-called wanted terms which are proportional to 
$\BB(\bt)$ and unwanted terms which are proportional to Bethe 
vectors with some of the Bethe parameters replaced by the 
parameters $z$ or $z_s$. We will prove that each unwanted 
term is labeled by a pair of  integers $(\ell,k)$ such that 
$n'\leq\ell < k\leq n$.

To describe the wanted and  unwanted terms 
for each $\ggo$ we introduce the operations $\Z^k_\ell$
for $n'\leq\ell \leq k\leq n$ of extension 
of the arguments of the Bethe vectors. We will use different extensions of the Bethe parameter sets:
\begin{equation}\label{exBCD1}
\bw^s=\{\bt^s,\bz_s\},\qquad \bu^s=\{\bt^s,z\},\qquad \bv^s=\{\bt^s,z_s\}\,.
\end{equation}

We define the $\Z^k_\ell$ operations uniformly for all $\ggo\not=\gln$
\begin{subequations}\label{Zop-all}
\begin{equation}\label{Zop-o21-sp}
\Z_\ell^k\cdot\BB(\bt)=\begin{cases}
\BB(\{\bw^s\}_0^{\ell'-1},\{\bu^s\}_{\ell'}^{k-1},\{\bt^s\}_{k}^{n-1}),
& \bvphi_\ggo\leq \ell'\leq k\leq n\,,\\[2mm]
\BB(\{\bw^s\}_0^{k-1},\{\bv^s\}_{k}^{\ell'-1},\{\bt^s\}_{\ell'}^{n-1}),
& \bvphi_\ggo\leq k\leq \ell'\leq n\,,\\[2mm]
\BB(\{\bt^s\}_0^{\ell-1},\{\bu^s\}_\ell^{k-1},\{\bt^s\}_k^{n-1}),
& \bvphi_\ggo\leq \ell\leq k\leq n\,,\\[2mm]
\BB(\{\bt^s\}_0^{k'-1},\{\bv^s\}_{k'}^{\ell'-1},\{\bt^s\}_{\ell'}^{n-1}),
 & \bvphi_\ggo\leq k'\leq\ell'\leq n\,.
\end{cases}
\end{equation}
In the case of $\ggo=\ggd$, there are extra definitions for the operations $\Z^{0,1}_{\ell}$, 
$-n+1\leq\ell\leq -1$ and the operations $\Z_{0,1}^{k}$, 
$2\leq k\leq n$  
\begin{equation}\label{Zop-o2nb}
\begin{split}
&\begin{array}{l}
\Z^0_{\ell}\cdot\BB(\bt)=
\BB(\bt^0,\bw^{1},\{\bv^s\}_2^{\ell'-1},\{\bt^s\}_{\ell'}^{n-1}),\\[2mm]
\Z^1_{\ell}\cdot\BB(\bt)=
\BB(\bw^0,\bt^{1},\{\bv^s\}_2^{\ell'-1},\{\bt^s\}_{\ell'}^{n-1}),
\end{array}
\quad
2\leq\ell'\leq n\,,\\[4mm]
&\begin{array}{l}
\Z^k_{1}\cdot\BB(\bt)=
\BB(\bt^0,\bw^{1},\{\bu^s\}_2^{k-1},\{\bt^s\}_k^{n-1}),\\[2mm]
\Z^k_{0}\cdot\BB(\bt)=
\BB(\bw^0,\bt^{1},\{\bu^s\}_2^{k-1},\{\bt^s\}_k^{n-1}),
\end{array}
\quad 2\leq k\leq n\,.
\end{split}
\end{equation}
\end{subequations}

Note that in all formulas \r{Zop-all} the operation 
$\Z^\ell_\ell\cdot\BB(\bt)=\BB(\bt)$ is the identity operation. 
Note also that for $\ggo=\mathfrak{o}_{2n}$ the operation 
$\Z^1_0$ is not defined. 

In the appendix~\ref{AppB} we introduce, for each algebra 
$\ggo$, functions $\psi_\ell(z;\bt)$, $\phi_\ell(z;\bt)$
for $n'\leq\ell\leq n$ and $\mu^k_\ell(z;\bt)$ for $n'\leq \ell\leq k\leq n$. 
Using these functions we formulate the following proposition which is a rewriting of  
proposition~\ref{ac-main} as a sum over partitions of the original 
sets of  Bethe parameters $\bt^s$:

\begin{prop}\label{diag-act}
The action of the diagonal monodromy entries $T_{i,i}(z)$ on the 
off-shell Bethe vectors $\BB(\bt)$ can be expressed as
\begin{equation}\label{d-ac1}
T_{i,i}(z)\cdot\BB(\bt)=\sum_{\ell=n'}^{i}\sum_{k=i}^n
\sum_{{\rm part}\ \bt}\, \lambda_k(z)\,
\mu^k_\ell(z;\bt_{\st})\,\Z^k_\ell\cdot \BB(\bt_{\st})\,
\Ups_i^{\ell,k}(z;\bt_{\so},\bt_{\st},\bt_{\sth})\,,
\end{equation}
where $n'\leq\ell\leq k\leq n$ and the functions 
$\Ups_i^{\ell,k}(z;\bt_{\so},\bt_{\st},\bt_{\sth})$ 
are defined by the equalities 
\begin{equation}\label{d-ac2}
\Ups_i^{\ell,k}(z;\bt_{\so},\bt_{\st},\bt_{\sth})=
\frac{\lambda_n(z)}{\lambda_k(z)}\,
\frac{\mu^n_{n'}(z;\bt)}{\mu^k_\ell(z;\bt_{\st})}\,
\frac{\coA_i(\bw_{\so})}{\psi_{n'}(z;\bw_{\so})}\, 
\frac{\roA_j(\bw_{\sth})}{\phi_n(z;\bw_{\sth})}\,
\Om(\bw_{\so},\bw_{\st},\bw_{\sth})\,  \alpha(\bw_{\sth})\,.
\end{equation}
In \r{d-ac1}, the cardinalities of the subsets in the partitions 
$\bt^s\dashv\{\bt^s_{\so},\bt^s_{\st},\bt^s_{\sth}\}$ 
are given in the appendix~\ref{AppA}. 
Partitions of the extended sets 
$\bw^s\dashv\{\bw^s_{\so},\bw^s_{\st},\bw^s_{\sth}\}$ appearing in the r.h.s. of \r{d-ac2}
can be reduced to the partitions $\bt^s\dashv\{\bt^s_{\so},\bt^s_{\st},\bt^s_{\sth}\}$, according to the 
description also given in appendix~\ref{AppA}. This reduction depends on the
integers $i$, $\ell$ and $k$ and the algebra $\ggo$.
\end{prop} 

\proof 
The proof of the proposition reduces to the description of the subsets 
in the partitions of the extended sets 
$\bw^s\dashv\{\bw^s_{\so},\bw^s_{\st},\bw^s_{\sth}\}$
which yields non-zero contribution  
in the action formula \r{T-act-all} for $i=j$. Then one has to substitute
these partitions in the formula \r{d-ac2} to get expressions for 
the functions $\Ups_i^{\ell,k}(z;\bt_{\so},\bt_{\st},\bt_{\sth})$. 
The fact 
that the partition are labeled by the integers $\ell$ and $k$ such 
that $n'\leq \ell\leq k\leq n$  is proved in the appendix~\ref{AppA}
together with the explicit description of the subsets 
$\bw^s_{\so}$, $\bw^s_{\st}$, and $\bw^s_{\sth}$.\qed 

\begin{cor}
The off-shell Bethe vectors have definite colors:
\begin{equation}
\col_{a}\, \BB(\bar t) = |\bar t^a|\, \BB(\bar t)\,, \quad a\in J_\ggo\,.
\end{equation}
\end{cor}
\proof
Considering the action formula \eqref{d-ac1} in the limit $z\to\infty$, the coefficient of $\frac1z$ provides the result.\qed


According to the proposition~\ref{diag-act} 
the action of the diagonal monodromy entries is given by the sum 
of the wanted terms when $\ell=k=i$
and the unwanted terms when $\ell<k$. It is shown 
in the appendix~\ref{AppA} that for the wanted terms 
the subsets $\bt^s_\so$ and $\bt^s_\sth$ are all empty and $\bt^s_\st=\bt^s$.
Moreover, using formulas \r{want1} 
one can verify that $\Upsilon_i^{i,i}(z;\vn,\bt,\vn)=1$
and the action of the 
transfer matrix \r{trans} on off-shell Bethe vectors $\BB(\bt)$ is given by the sum 
\begin{equation}\label{tr-ac}
\begin{split}
\mathcal{T}_\ggo(z)\cdot \BB(\bar t)=\ &
\sum_{i=n'}^n
\sum_{\ell=n'}^{i}\sum_{k=i}^n 
\sum_{{\rm part}\ \bt} \lambda_k(z)\,\mu_{\ell}^k(z;\bt_{\st})\,\Z_\ell^k\cdot\BB(\bt_{\st})
 \Ups_{i}^{\ell,k}(z;\bt_{\so},\bt_{\st},\bt_{\sth})=\\
 =\ &\BB(\bt)\,\sum_{i=n'}^n\lambda_i(z)\,
\mu_i^i(z;\bt)+\\
&+\sum_{\ell=n'}^{n-1}\sum_{k=\ell+1}^n 
\sum_{{\rm part}\ \bt} \lambda_k(z)\,\mu_{\ell}^k(z;\bt_{\st})\,\Z_\ell^k\cdot\BB(\bt_{\st})
\sum_{i=\ell}^k \Ups_{i}^{\ell,k}(z;\bt_{\so},\bt_{\st},\bt_{\sth})\,.
\end{split}
\end{equation}
The possibility to exchange 
the order of summation in \r{tr-ac} follows from the fact that the subsets 
$\bt^s_{\st}$ does not depend on $i$ for  $\ell<k$.
Note also that for $\mathfrak{o}_{2n}$-invariant model 
the term in the second line of \r{tr-ac} with $\ell'=k=1$ is absent
because the function 
$\mu^1_0(z;\bt)$ given by \r{mu-o2na} vanishes in that case. 

\begin{prop}\label{eigen}
If the set of  Bethe parameters satisfy the Bethe equations 
\r{BE-all} we get 
\begin{equation}\label{eig1}
\sum_{i=\ell}^k \Ups_{i}^{\ell,k}(z;\bt_{\so},\bt_{\st},\bt_{\sth})  = 0
\end{equation}
 for all pairs $(\ell,k)$ such that 
$n'\leq \ell< k\leq n$. Then, the on-shell Bethe vector $\BB(\bt)$ becomes 
an eigenvector of transfer matrix $\mathcal{T}_\ggo(z)$ with eigenvalue
\begin{equation*}\label{eig2}
\tau_\ggo(z;\bt)=\sum_{i=n'}^n \lambda_i(z)\,\mu_i^i(z;\bt). 
\end{equation*}
\end{prop}
\proof This proposition is proved by a direct verification of the 
vanishing condition \r{eig1} using \r{d-ac2} and the explicit description of the 
subsets given in the appendix~\ref{AppA}.
\qed

We want to stress that the proposition~\ref{eigen} not only verifies  the 
eigenvector property of the on-shell Bethe vectors, but provides an explicit 
description of all wanted and unwanted terms in the action of transfer 
matrix on the off-shell Bethe vectors in arbitrary $\ggo$-invariant 
integrable model. 

The eigenvalues $\tau_\ggo(z;\bt)$ for each of the algebra $\ggo$ 
can be explicitly written using functions 
which are defined by  \r{g-h-f-gen}.

\begin{itemize}

\item For $\ggo=\mathfrak{o}_{2n+1}$. 
The eigenvalue of the transfer matrix is equal to
\begin{equation}\label{BEB3}
\begin{split}
\tau_{\mathfrak{o}_{2n+1}}(z;\bt)&=\lambda_0(z)\, f(\bt^0,z_0)\, f(z,\bt^0) +\\
&+\sum_{i=1}^n \Big(\lambda_i(z)\, f(\bt^i,z)\, f(z,\bt^{i-1})
+\lambda_{i'}(z)\, f(\bt^{i-1},z_{i-1})\, f(z_{i},\bt^{i})\Big)\,,
\end{split}
\end{equation}
where we recall that $z_i=z-c(i-1/2)$ for $i=0,1,\ldots,n$ 
and the functional parameters $\lambda_{i'}(z)$ for 
$0\leq i\leq n$ are defined by the equalities 
\begin{equation}\label{sym-diag}
\lambda_{i'}(z)=\frac{1}{\lambda_i(z_i)}\ \prod_{s=i+1}^n
\frac{\lambda_s(z_{s-1})}{\lambda_s(z_s)}\,.
\end{equation}

\item For $\ggo=\mathfrak{sp}_{2n}$. 
The eigenvalue of the transfer matrix is equal to
\begin{equation}\label{BEC3}
\tau_{\mathfrak{sp}_{2n}}(z;\bt)=\
\sum_{i=1}^n \Big(\lambda_i(z)\ f(\bt^i,z)\ f_{i-1}(z,\bt^{i-1})
+\lambda_{i'}(z)\ f_{i-1}(\bt^{i-1},z_{i-1})\ 
f(z_{i},\bt^{i})\Big)\,,
\end{equation}
where the functional parameters $\lambda_{i'}(z)$ for 
$1\leq i\leq n$ are defined by the equalities \r{sym-diag} and 
we recall that $z_0=z$ and $z_i=z-c(i+1)$ for $i=1,\ldots,n-1$.

\item For $\ggo=\mathfrak{o}_{2n}$.
The eigenvalue of the transfer matrix is equal to 
\begin{equation}\label{BED2}
\begin{split}
\eigen_{\mathfrak{o}_{2n}}(z;\bt)&=\sum_{i=1}^n\left(
\lambda_i(z)\,f(z,\bt^{i'}) \, f(\bt^i,z)\, \prod_{q=1}^{\nus_{i-2}}f(z,\bt^{i-q})+\right.\\
&\qquad\qquad\qquad\qquad
+\left.\lambda_{i'}(z)\,f(\bt^{i'},z)\,f(z_i,\bt^{i})\,\prod_{q=1}^{\nus_{i-2}}
f(\bt^{i-q},z_{i-q})\right)\,,
\end{split}
\end{equation}  
where the sets $\bt^s=\vn$ for $s<0$ and 
we set $\nu_{-1}=0$. We recall that 
$z_0=z_1=z$ and $z_i=z-c(i-1)$ for $i=2,\ldots,n-1$.
The functional parameters $\lambda_{i'}(z)$ for 
$1\leq i\leq n$ are defined by the equalities \r{sym-diag}.

One
we can rewrite \eqref{BED2} as
\begin{equation}\label{BED3}
\begin{split}
\eigen_{\mathfrak{o}_{2n}}(z;\bt)=\ &\lambda_2(z)^{-1}\ f(\bt^2,z)^{-1}
\prod_{q=0}^1\Big(f(\bt^q,z)\,\lambda_q(z)+f(z,\bt^q)\,f(\bt^2,z)\,\lambda_2(z)\Big)
+\\
&+\sum_{i=3}^n \Big(\lambda_i(z)\ f(\bt^i,z)\ f(z,\bt^{i-1})
+\lambda_{-i+1}(z)\ f(\bt^{i-1},z_{i-1})\ f(z_i,\bt^{i})\Big)\,.
\end{split}
\end{equation}
To show the equality between the two expressions
\r{BED2} and  \r{BED3}, we used the relation $f(\bt^2,z)=f(z_2,\bt^2)^{-1}
=f(z-c,\bt^2)^{-1}$ and 
\begin{equation*}\label{lamrel}
\lambda_{-1}(z)=\frac{\lambda_0(z)\,\lambda_1(z)}{\lambda_2(z)}
\end{equation*}
which is a direct consequence of \r{sym-diag}.
\end{itemize} 

For $n=2$ the existence of such factorized presentation of the eigenvalue 
\r{BED3} is a consequence of embedding $\mathfrak{o}_4$-invariant 
monodromy matrix into product of two $\mathfrak{gl}_2$-invariant 
monodromies described in the paper \cite{AMR}.

\subsection{Rectangular recurrence relations for off-shell Bethe vectors\label{sec:rec-rec}}

For $n'\leq i\leq\ell\leq n-1$ and 
$n'+1\leq k\leq j\leq n$ let
$\sPh^\ell_i(z;\bt_{\so})$ and $\oPh^k_j(z;\bt_{\sth})$ be the functions 
defined  by the equalities 
\begin{itemize}
\begin{subequations}\label{oP-all}

\item for $\ggo=\mathfrak{o}_{N}$:
\begin{equation}\label{oPoN}
\sPh^\ell_i(z;\bt_{\so})=(\sigma_{i'-\frd_\ggo})^{\delta_{\ell\geq\frd_\ggo}},\qquad
\oPh^k_j(z;\bt_{\sth})=(\sigma_j)^{\delta_{k\leq-\frd_\ggo}}\,,
\end{equation}

\item for $\ggo=\mathfrak{sp}_{2n}$:
\begin{equation}\label{oPsp}
\begin{split}
\sPh^\ell_i(z;\bt_{\so})&=
\begin{cases}1,&-n+1\leq\ell\leq 1\,,\\
\ophf_\ell(\bup_\so),&2\leq\ell\leq n-1\,,
\end{cases}\\
\oPh^k_j(z;\bt_{\sth})&=
\begin{cases}
    \ophf_{k'}(\bvp_{\sth}),&-n+2\leq k\leq -1\,,\\
    1,&0\leq k\leq n\,,
\end{cases}
\end{split}
\end{equation}
where the sign factors $\sigma_m$ are defined in \eqref{sigma} and the functions 
$\ophf_\ell(\bup_{\so})$, $\ophf_{k'}(\bvp_{\sth})$ 
are given by \r{phf-sp} for the subsets $\bup_\so$
and $\bvp_\sth$ described below.

\end{subequations}
\end{itemize}

In \r{oPsp} the functions $\sPh^\ell_i(z;\bt_{\so})$ and $\oPh^k_j(z;\bt_{\sth})$
depend on the indices $i$ and $j$ implicitly through the cardinalities of 
the subsets $\bt_{\so}$ and $\bt^s_{\sth}$ given by \r{oC-g}. 
The subsets $\bup^s_{\so}$ and $\bvp^s_{\sth}$
in the arguments of the functions 
$\ophf_\ell(\bup_{\so})$ and $\ophf_{k'}(\bvp_{\sth})$
are  
\begin{equation*}
\bup^s_{\so}=\begin{cases}
\{z,\bt^\ell_{\so}\},&s=\ell\,,\\
\bt^s_\so,&1\leq s\leq \ell-1\,,
\end{cases}\qquad 
\bvp^s_{\sth}=\begin{cases}
\{z_{k'},\bt^{k'}_{\sth}\},&s=k'\,,\\
\bt^s_\sth,&1\leq s\leq k'-1\,.
\end{cases}
\end{equation*}

For each algebra $\ggo$ we define the $\ggo$-dependent
absolute value of the integer $\ell$ by the rule 
\begin{equation*}\label{abs-v}
|\ell|_\ggo=|\ell-\frd_\ggo/2|+\frd_\ggo/2\,,
\end{equation*}
where $|\ell|$ is the usual absolute value. A significant property of 
the $\ggo$-dependent absolute value $|\ell|_\ggo$ 
is that $|\ell'|_\ggo=|\ell|_\ggo$ for all  for $\ggo\not=\mathfrak{gl}_{n}$, 
which means, 
in particular, that 
for $\ggo=\mathfrak{sp}_{2n}$ and  $\mathfrak{o}_{2n}$ we have 
$|0|_\ggo=|1|_\ggo=1$. For $\ggo=\gln$ and  $\ggo=\ggb$, $|\ell|_\ggo$ is just the usual absolute value.

Using the definition \r{oP-all} of the functions 
$\sPh^\ell_i(z;\bt_{\so})$ and $\oPh^k_j(z;\bt_{\sth})$
we can formulate the following proposition. 

\begin{prop}\label{rec-rr}
For any integers $\ell$ and $k$ such that 
$n'\leq \ell< k\leq n$ 
the off-shell Bethe vectors in $\ggo$-invariant integrable models 
satisfy the rectangular recurrence relations  
\begin{equation}\label{rr1}
\Z^k_\ell\cdot\BB(\bt)=\frac{1}{\lambda_k(z)\,\mu^k_\ell(z;\bt)}\ 
\sum_{i=n'}^\ell\sum_{j=k}^n\sum_{{\rm part}\ \bt}
\alpha(\bt_{\sth})\,
\Xi^{\ell,k}_{i,j}(z;\bt_{\so},\bt_{\st},\bt_{\sth})\,T_{i,j}(z)\cdot\BB(\bt_{\st})\,,
\end{equation}
where $\Z^k_\ell$ act by \eqref{Zop-all}, and 
\begin{equation}\label{Xi-rr}
\Xi^{\ell,k}_{i,j}(z;\bt_{\so},\bt_{\st},\bt_{\sth})=
\sPh^\ell_i(z;\bt_{\so})\,\oPh^k_j(z;\bt_{\sth})\,\psi_\ell(z;\bt_{\so})\,
\phi_k(z;\bt_{\sth})\,\Om(\bt_{\so},\bt_{\st})\,
\Om(\bt_{\st},\bt_{\sth})
\end{equation}
and the sum in \r{rr1} goes over partitions of the sets
$\bt^s\dashv\{\bt^s_{\so},\bt^s_{\st},\bt^s_{\sth}\}$ with cardinalities 
\begin{subequations}\label{oC-g}
 \begin{eqnarray}
|\bt^s_{\so}|&=&\begin{cases}
\Theta(\ell-\bvphi_\ggo)
\Theta(i'-\bvphi_\ggo)+
\Big(\Theta(\ell-\bvphi_\ggo)\,\delta_{s,i}
+\Theta(i'-\bvphi_\ggo)\delta_{s,\ell'}\Big)\delta_{\bvphi_\ggo,2}
,\quad 0\leq s< \bvphi_\ggo\,,\\
\Theta(\ell-\bvphi_\ggo)
\Big(\Theta(i'+s-\xi_\ggo)+\Theta(i'-s-1)\Big)\,,
\quad \bvphi_\ggo \leq s<|\ell|_\ggo,\\
\Theta(i'-s-1)\,,
\quad |\ell|_\ggo+\delta_{\bvphi_\ggo,2}\leq s\leq n-1\,,
\end{cases}\label{oCg-tI}\\
|\bt^s_{\sth}|&=&\begin{cases}
\Theta(k'-\bvphi_\ggo)
\Theta(j-\bvphi_\ggo)+
\Big(\Theta(k'-\bvphi_\ggo)\delta_{s,j'}
+\Theta(j-\bvphi_\ggo)\delta_{s,k}\Big)\delta_{\bvphi_\ggo,2}
\,,\quad 0\leq s< \bvphi_\ggo,\\
\Theta(k'-\bvphi_\ggo)
\Big(\Theta(j+s-\xi_\ggo)+\Theta(j-s-1)\Big)\,,
\quad \bvphi_\ggo \leq s<|k|_\ggo,\\
\Theta(j-s-1),
\quad |k|_\ggo+\delta_{\bvphi_\ggo,2}\leq s\leq n-1.
\end{cases}
\label{oCg-tII}
\end{eqnarray}
\end{subequations}
\end{prop}

Note that first lines in the description of the cardinalities 
$|\bt^s_\so|$ and $\bt^s_\sth|$ \r{oC-g} are absent for $\ggo=\ggb$ (see \r{oCo2n1}).
For $\ggo=\gsp$ these lines describe the cardinalities of the subsets
$\bt^0_\so$ and $\bt^0_\sth$ only and they can be unified with the second 
lines of \r{oCg-tI} and \r{oCg-tII} (see \r{oCsp}).

One can see from the formulas \r{oC-g} that the cardinalities of the subsets 
$\bt^s_{\so}$ are defined solely by the index $i$, while
the cardinalities of $\bt^s_{\sth}$ are defined solely by the index $j$.
Moreover, the 
cardinalities of the subsets $\bt^s_{\sth}$ can be obtained 
from the cardinalities $\bt^s_{\so}$ by the formal replacement 
$i'\to j$ and $\ell\to k'$. 
The second line in \r{oCg-tI} and \r{oCg-tII} are 
 absent when $\ell=k=0$ for 
$\ggo=\mathfrak{sp}_{2n}$ and $\ell=k=0,1$ for 
$\ggo=\mathfrak{o}_{2n}$ since in these cases the inequalities
$\bvphi_\ggo\leq s<|\ell|_\ggo$ and 
$\bvphi_\ggo\leq s<|k|_\ggo$ cannot be fulfilled for any $s$. 
In these cases the sums $|\ell|_\ggo+\delta_{\bvphi_\ggo,2}$
and $|k|_\ggo+\delta_{\bvphi_\ggo,2}$ becomes 1 for 
$\ggo=\mathfrak{sp}_{2n}$ and 2 for $\ggo=\mathfrak{o}_{2n}$.

\proof Note first that for $\ell'=k=n$ there is no summation 
over $i$ and $j$ 
in the recurrent relation \r{rr1}. The summation over partitions is also absent since 
for the values $\ell'=k=n$ the cardinalities 
of all subsets $|\bt^s_{\so}|=|\bt_{\sth}|=0$,  according to \r{oC-all}. The recurrence relation becomes 
equivalent to the action of the highest monodromy entry \r{T-act-g}. 
Applying to this relation the appropriate zero mode operator, using 
the relation~\r{bvde1} and the commutation relations \r{zT-all}
we prove by induction the rectangular recurrence relation \r{rr1} 
for all $\ell$ and $k$ such that $n'\leq \ell<k\leq n$.  
\qed
We call these 
recurrence relations rectangular because in their right hand side 
 appear only  monodromy entries from the upper-right rectangular domain of 
the monodromy matrix defined by the integers $\ell$ and $k$.

Among the recurrence relations, the basic ones occur when 
$k=\ell+1$ for $\ell\in\PSR_\ggo$. They describe the extension 
of the set of  Bethe parameters $\bt^\ell$ by a single parameter $z$. 
These basic recurrence relations can be seen as building blocks that
allow to construct any rectangular recurrence relation. 
They are however the last ones that one obtains starting from the action of the highest monodromy entry.
They also allow to build 
recursively an arbitrary 
off-shell Bethe vector starting from the vacuum vector 
 proving in this way its uniqueness, up to a re-ordering of the terms using  the commutation relation \r{rtt}. An explicit form for these basic recurrence relations can be found
 in the paper \cite{LPR25} in the case of the algebra $\ggo=\ggb$.

The cardinalities of the subsets  
 used in the sum 
over partitions in \r{rr1} can be explictly written 
for different $\ggo$ as follows.

\begin{itemize}
\begin{subequations}\label{oC-all}

\item For $\ggo=\mathfrak{o}_{2n+1}$:
\begin{equation}\label{oCo2n1}
\begin{split}
|\bt^s_{\so}|&=\begin{cases}
\displaystyle \Theta(\ell)
\Big(\Theta(s-i)+\Theta(-s-i-1)\Big)\,,& \displaystyle
 s< |\ell|\,,\\[2mm]
\Theta(-s-i-1)\,,& \displaystyle 
s\geq|\ell|\,,
\end{cases}\\[3mm]
|\bt^s_{\sth}|&=\begin{cases}
\displaystyle \Theta(-k)
\Big(\Theta(j+s)+\Theta(j-s-1)\Big)\,,& \displaystyle
 s< |k|\,,\\[2mm]
\Theta(j-s-1)\,,& \displaystyle 
s\geq |k|\,.
\end{cases}
\end{split}
\end{equation}

\item For $\ggo=\mathfrak{sp}_{2n}$:
\begin{equation}\label{oCsp}
\begin{split}
|\bt^s_{\so}|&=\begin{cases}
\displaystyle \frac{\Theta(\ell-1)}{1+\delta_{s,0}}
\Big(\Theta(s-i)+\Theta(-s-i)\Big)\,,& \displaystyle
0\leq s< \Big|\ell-\frac12\Big|+\frac12\,,\\[2mm]
\Theta(-s-i)\,,& \displaystyle 
\Big|\ell-\frac12\Big|+\frac12\leq s\leq n-1\,,
\end{cases}\\[3mm]
|\bt^s_{\sth}|&=\begin{cases}
\displaystyle \frac{\Theta(-k)}{1+\delta_{s,0}}
\Big(\Theta(j+s-1)+\Theta(j-s-1)\Big)\,,& \displaystyle
0\leq s< \Big|k-\frac12\Big|+\frac12\,,\\[2mm]
\Theta(j-s-1)\,,& \displaystyle 
\Big| k-\frac12\Big|+\frac12\leq s\leq n-1\,.
\end{cases}
\end{split}
\end{equation}

\item For $\ggo=\mathfrak{o}_{2n}$:
\begin{equation}\label{oCo2n}
\begin{split}
|\bt^s_{\so}|&=\begin{cases}
\Theta(\ell-2)\Big(\Theta(-i-1)+\delta_{i,s}\Big)+\Theta(-i-1)\,\delta_{s,\ell'}\,
\delta_{\ell=0,1}\,,\quad s=0,1\,,\\
\Theta(\ell-2)\Big(\Theta(s-i)+\Theta(-s-i)\Big)\,,\quad 2\leq s<\Big|\ell-\frac12\Big|+
\frac12\,,\\
\Theta(-i-s)\,,\quad \Big|\ell-\frac12\Big|+\frac12+\delta_{\ell=0,1}\leq s\leq n-1\,,
\end{cases}\\[3mm]
|\bt^s_{\sth}|&=\begin{cases}
\Theta(-k-1)\Big(\Theta(j-2)+\delta_{j,s'}\Big)+\Theta(j-2)\,\delta_{s,k}\,
\delta_{k=0,1}\,,\quad s=0,1\,,\\
\Theta(-k-1)\Big(\Theta(s+j-1)+\Theta(-s+j-1)\Big)\,,
\quad 2\leq s< \Big|k-\frac12\Big|+\frac12\,,\\
\Theta(-s+j-1)\,,\quad \Big|k-\frac12\Big|+\frac12+\delta_{k=0,1}\leq s\leq n-1\,.
\end{cases}
\end{split}
\end{equation}

\end{subequations}
\end{itemize}

\section{Coproduct properties of off-shell Bethe vectors}\label{sect6}

The coproduct properties of $\mathfrak{gl}_2$ type Bethe vectors 
were described for the first time in the paper \cite{IK84} on an example 
of generalized two-site model. In the book \cite{S22} 
this type of integrable models were called {\it composite}. 
 
To describe a composite model we consider two generic $\ggo$-invariant 
 integrable models which are described by two monodromy matrices 
 $T^{[1]}(u)$ and $T^{[2]}(u)$ satisfying the $RTT$ commutation relation 
 with the same $\ggo$-invariant $R$-matrix. It means, in particular, that 
 they are matrices of the same size  and one can consider 
 the monodromy matrix of the composed model with the matrix entries 
 \begin{equation}\label{comp-m}
 T^{[2,1]}_{i,j}(u)=\sum_{\ell=n'}^n T^{[2]}_{i,\ell}(u)\,T^{[1]}_{\ell,j}(u)\,,
 \end{equation}
 given by   the matrix multiplication in the auxiliary 
 space which is the same for all three models. Denote by 
 $\BB^{[2]}(\bt)$ and $\BB^{[1]}(\bt)$ the off-shell Bethe vectors 
 built from the upper-triangular monodromy matrix entries 
 $T^{[2]}_{i,j}(u)$, $T^{[1]}_{i,j}(u)$, $i<j$  using the vacuum vectors 
 $\rvac^{[2]}$ and  $\rvac^{[1]}$ respectively. 
 The vacuum vector $\rvac$ in the composite model is defined as 
 $\rvac^{[2]}\otimes\rvac^{[1]}$.
 
 The coproduct properties of the off-shell Bethe vectors describe a possibility 
 to built the Bethe vectors in a composite model through the Bethe vectors 
 of their component. The following propositions demonstrate that these 
 properties are consequence of the defining relations for off-shell Bethe 
 vectors in a generic $\ggo$-invariant integrable model.

 Let us define disjoint partitions $\bt^s\dashv\{\bt^s_\so,\bt^s_\st\}$ 
 such that the cardinalities of the subsets 
 $\bt^s_\so$ and $\bt^s_\st$ are $0\leq |\bt^s_\so|\leq |\bt^s|$ and 
 $|\bt^s_\st|=|\bt^s|-|\bt^s_\so|$. 
 Let  $\Om(\bt_\so,\bt_\st)$ be  the function defined by  
 \begin{equation}\label{Om-gen}
\Om(\bt_{\so},\bt_{\st})=\prod_{a\,\in\,\PSR_{\ggo}}\Omr_a(\bt_{\so},\bt_{\st})
=\prod_{a\,\in\,\PSR_{\ggo}}\Oml_a(\bt_{\so},\bt_{\st})=
\prod_{a,b\,\in\,\PSR_{\ggo}}\bgam^b_a(\bt^a_{\so},\bt^b_{\st})\,.
\end{equation}

 \begin{prop}\label{copr-pr-IK}
 As result of the defining relation~\ref{def:BV} for the off-shell Bethe 
 vectors $\BB^{[2]}(\bt)$ and $\BB^{[1]}(\bt)$, 
 the composed Bethe vector given by 
 \begin{equation}\label{cop1}
 \BB^{[2,1]}(\bt)=\sum_{\rm part}\Om(\bt_\so,\bt_\st)\,
 \BB^{[2]}(\bt_\so)\,\BB^{[1]}(\bt_\st)\ \prod_{s\in\PSR_\ggo}
 \alpha_s^{[2]}(\bt^s_\st)\,,
 \end{equation}
 also obeys the definition~\ref{def:BV}.
 The function $\Om(\bt_\so,\bt_\st)$ is defined in \eqref{Om-gen}
  \end{prop}
 \proof
We sketch the proof of this proposition by a combined induction 
over the cardinalities of the subsets and the rank of the algebra $\ggo$.  

We first consider $\ggo=\gln$.
First of all, for $\ggo=\mathfrak{gl}_2$, the relation \r{cop1} was proved in \cite{IK84} 
(see proposition~4.1 in the book
\cite{S22}). Then using the embedding of $\gl{2}$ in $\gl{3}$  
and the recurrence relation given by proposition~\ref{rr-gln} for 
$\gl{3}$-type off-shell Bethe vectors, we can prove \r{cop1} 
for the Bethe vector $\BB^{[2,1]}(\bt^1,\{z\})$. 
 Using again the recurrence relation on this Bethe vector,
 we prove by induction the relation \r{cop1} for an arbitrary cardinality of the 
set $\bt^2$ in the composed Bethe vector  $\BB^{[2,1]}(\bt^1,\bt^2)$ for $\ggo=\mathfrak{gl}_3$.
Continuing we can prove that \r{cop1} is valid for $\gl{n-1}$ 
composed Bethe vectors $\BB^{[2,1]}(\{\bt^s\}_1^{n-2})$. 
At the last step of the induction on the rank we first prove 
\r{cop1} for the $\gl{n}$ type Bethe vector $\BB^{[2,1]}(\{\bt^s\}_1^{n-2},\{z\})$
and then using induction over the cardinalities of the set $\bt^{n-1}$ 
we finally prove statement of the proposition.

For the other algebras $\ggo=\ggb$, $\gsp$ and $\ggd$, we use 
the embedding of $\mathfrak{gl}_{n}$ in these algebras.
Since \r{cop1} has been obtained for these cases, we use the $\mathfrak{gl}_{n}$ Bethe vectors as
 the starting point for the induction on the rank of $\ggo$. Then, the proof goes along the same lines as for $\gln$.
The fact that $\mathfrak{gl}_{n}$ Bethe vectors are a subset of the $\ggo$ Bethe vectors has been proven for $\ggo=\ggb$ in the paper \cite{LPR25}.
The cases $\ggo=\gsp$ and $\ggo=\ggd$ can de dealt analogously.
 \qed

 The property \r{cop1}  of the composed Bethe vectors was called a
 coproduct property because if a generic monodromy matrix 
 is realized as fundamental $T$-operator of the Yangian $Y(\ggo)$ 
 then the formulas \r{comp-m} are nothing but the coproduct map $\Delta$
 of the fundamental $T$-operator 
 \begin{equation*}\label{cop3}
 \Delta\,T_{i,j}(u)=\sum_{\ell=n'}^n T_{\ell,j}(u)\otimes T_{i,\ell}(u)\,.
 \end{equation*}

\section{Conclusion and discussion}
In this paper, we have introduced a unified and intrinsic definition of off-shell Bethe vectors formulated in terms of two fundamental relations. Remarkably, these relations suffice to reconstruct the full algebraic structure underlying the Bethe ansatz in the RTT presentation of the Yangian 
$Y(\ggo)$, thereby providing a minimal and conceptually transparent framework for their construction.

This approach sheds new light on the relationship between different realizations of Yangian symmetry. In particular, it paves the way for a systematic derivation of Bethe vectors from the projection method, establishing a deep and explicit correspondence between the $RTT$ formulation and the 
Drinfeld's 'new' realization of the Yangian double  $DY(\ggo)$. A complete proof of this correspondence will be presented in \cite{LPR-PM}. 

The proposed definition naturally invites further structural investigations. In particular, morphisms between Yangians associated with Lie algebras of small rank are expected to induce nontrivial and computable relations between the corresponding Bethe vectors, whose explicit construction is the subject of our forthcoming article \cite{LPR-SR}.

Beyond classical series, our framework offers a promising avenue for the study of exceptional Lie algebras 
$G_2$, $F_4$, and $E_n$ with 
$n=6,7,8$,  where analogous questions remain largely unexplored.
Clearly, extensions to quantum affine algebras can also be dealt within this framework.

As a side remark, we have introduced two parameters, $\epsilon_\ggo$ and $\xi_\ggo$, which allow us to treat all the algebras $\ggo$ in a uniform manner. Notice that $\epsilon_\ggo$ was already used in the context of twisted Yangians to provide a uniform presentation thereof, see \cite{MNO} where it appeared as a sign $\pm$ for $\mathfrak{o}_N$ and $\mathfrak{sp}_{2n}$. A natural question then arises as to whether these parameters admit a more intrinsic algebraic interpretation, for instance in terms of roots and/or weights of the corresponding algebras.

More broadly, the present results lay a solid foundation for future applications, notably to the computation of scalar products, form factors, and correlation functions in integrable models with higher-rank symmetry.

\section*{Acknowledgement}
S.P. acknowledges support from the PAUSE Programme (2024020408) and the hospitality of LAPTh, where this work was conducted.
The work of A.~L. was supported by the Beijing Natural Science Foundation (IS24006) and Beijing
Talent Program.

\appendix

\section{Collections of  rational functions}\label{AppB}

In this appendix we describe for each algebra $\ggo$ 
 the functions which are used through the paper. 

For any $\ell$ such that $n'\leq\ell\leq n$
we introduce two functions $\psi_\ell(z;\bt)$ and $\phi_\ell(z;\bt)$ which depend on $\ggo$. We write 
their definitions explicitly for each $\ggo$.
\begin{itemize}
\begin{subequations}\label{ps-ph-all}

\item For $\ggo=\mathfrak{o}_{2n+1}$:
\begin{equation}\label{ps-ph-o2n1}
\begin{split}
\psi_\ell(z;\bt)&=\begin{cases}
\displaystyle
h(\bt^\ell,z)\, g(z,\bt^{\ell-1}),
& 1\leq \ell\leq n\,,\\[2mm]
\displaystyle
g(z_0,\bt^{0}),
& \ell=0\,,\\[2mm]
\displaystyle
(-1)^{|\bt^{\ell'}|+|\bt^{\ell'-1}|}\,g(z_{\ell'},\bt^{\ell'})\,
h(\bt^{\ell'-1},z_{\ell'-1}),
& 1\leq \ell'\leq n\,,
\end{cases}\\[6mm]
\phi_\ell(z;\bt)&=\begin{cases}
\displaystyle
g(\bt^\ell,z)\, h(z,\bt^{\ell-1}),
& 1\leq \ell\leq n\,,\\[2mm]
\displaystyle
g(z,\bt^{0}),
& \ell=0\,,\\[2mm]
\displaystyle
(-1)^{|\bt^{\ell'}|+|\bt^{\ell'-1}|}\,h(z_{\ell'},\bt^{\ell'})\,
g(\bt^{\ell'-1},z_{\ell'-1}),
& 1\leq \ell'\leq n\,,
\end{cases}
\end{split}
\end{equation}

\item For $\ggo=\mathfrak{sp}_{2n}$:
\begin{equation}\label{ps-ph-sp}
\begin{split}
\psi_\ell(z;\bt)&=\begin{cases}
\displaystyle
h(\bt^\ell,z)\, g(z,\bt^{\ell-1}),
& 1\leq \ell\leq n\,,\\[2mm]
\displaystyle
(-1)^{|\bt^{\ell'}|+|\bt^{\ell'-1}|}\,g(z_{\ell'},\bt^{\ell'})\,
h_{\ell'-1}(\bt^{\ell'-1},z_{\ell'-1}),
& 1\leq \ell'\leq n\,,
\end{cases}\\[6mm]
\phi_\ell(z;\bt)&=\begin{cases}
\displaystyle
g(\bt^\ell,z)\, h_{\ell-1}(z,\bt^{\ell-1}),
& 1\leq \ell\leq n\,,\\[2mm]
\displaystyle
(-1)^{|\bt^{\ell'}|+|\bt^{\ell'-1}|}\,h(z_{\ell'},\bt^{\ell'})\,
g(\bt^{\ell'-1},z_{\ell'-1}),
& 1\leq \ell'\leq n\,.
\end{cases}
\end{split}
\end{equation}

\item For $\ggo=\mathfrak{o}_{2n}$:
\begin{equation}\label{ps-o2n}
\psi_\ell(z;\bt)=\begin{cases}
\displaystyle
h(\bt^\ell,z)\,\prod_{q=1}^{\nus_{\ell-2}}g(z,\bt^{\ell-q}),
& 2\leq \ell\leq n\,,\\[6mm]
\displaystyle
(-1)^{|\bt^{0}|+|\bt^{1}|}\, g(z_{\ell'},\bt^{\ell'})\,h(\bt^\ell,z_\ell),
& 0\leq \ell\leq 1\,,\\[4mm]
\displaystyle
(-1)^{|\bt^{\ell'}|}g(z_{\ell'},\bt^{\ell'})\,
\prod_{q=1}^{\nus_{\ell'-2}} (-1)^{|\bt^{\ell'-q}|}
h(\bt^{\ell'-q},z_{\ell'-q}),
& 2\leq \ell'\leq n\,,
\end{cases}
\end{equation}
\begin{equation}\label{ps-ph-o2n}
\phi_\ell(z;\bt)=\begin{cases}
\displaystyle
g(\bt^\ell,z)\,\prod_{q=1}^{\nus_{\ell-2}}h(z,\bt^{\ell-q}),
& 2\leq \ell\leq n\,,\\[6mm]
\displaystyle
 h(z_{\ell'},\bt^{\ell'})\,g(\bt^\ell,z_\ell),
& 0\leq \ell\leq 1\,,\\[4mm]
\displaystyle
(-1)^{|\bt^{\ell'}|}h(z_{\ell'},\bt^{\ell'})\,
\prod_{q=1}^{\nus_{\ell'-2}} (-1)^{|\bt^{\ell'-q}|}
g(\bt^{\ell'-q},z_{\ell'-q}),
& 2\leq \ell'\leq n\,.
\end{cases}
\end{equation}
\end{subequations}
\end{itemize}

Recall that the shifted parameters $z_{\ell'}$ in the formulas 
\r{ps-ph-all} have an expression which depends on $\ggo$ according 
to the definitions \r{z-sh1}. Note also that in most cases 
the function $\phi_\ell(z;\bt)$ can be obtained from the function 
$\psi_\ell(z;\bt)$ by replacing everywhere the function 
$g(u,v)$ by $h(u,v)$ and vice versa. 
However, this rule does not apply when one consider 
the special simple roots in $\ggo$
which are not of $\gln$ type.

Using the functions \r{ps-ph-all}, we introduce 
the functions $\mu^k_\ell(z;\bt)$ for $n'\leq \ell\leq k\leq n$ 
occurring in the eigenvalue property of on-shell Bethe vectors and in the
construction of the rectangular recurrence relations for off-shell Bethe vectors.
We  explicitly write them  for each algebra $\ggo$.

\begin{itemize}
\begin{subequations}\label{mu-all}

\item For $\ggo=\mathfrak{o}_{2n+1}$:
\begin{equation}\label{mu-o2n1a}
\begin{split}
\mu^k_\ell(z;\bt)&=\sigma_{\ell'-k}\, 
\Big(\frac{g(z_1,\bt^0)}{h(z,\bt^0)}\Big)^{\delta_{\ell<0}\,\delta_{k>1}}\ 
(k-1/2)^{\delta_{k,\ell'}}
\ \psi_\ell(z;\bt)\ \phi_k(z;\bt),
\quad \ell<k\,,
\end{split}
\end{equation}
\begin{equation}\label{mu-o2n1b}
\begin{split}
\mu^\ell_\ell(z;\bt)&=\Big(\frac{h(z,\bt^0)}{g(z_1,\bt^0)}\Big)^{\delta_{\ell,0}}
 \ \psi_\ell(z;\bt)\ \phi_\ell(z;\bt)=\\
&= \begin{cases}
f(\bt^\ell,z)\,f(z,\bt^{\ell-1})\,,&0<\ell\leq n\,,\\
f(\bt^0,z_0)\,f(z,\bt^{0})\,,&\ell=0\,,\\
f(\bt^{-\ell-1},z_{-\ell-1})\,f(z_{-\ell},\bt^{-\ell})\,,&-n\leq\ell<0\,.\\
\end{cases}\,.
\end{split}
\end{equation}

\item For $\ggo=\mathfrak{sp}_{2n}$:
\begin{equation}\label{mu-spa}
\begin{split}
\mu^k_\ell(z;\bt)&=\sigma_{\ell'-k}\, \sigma_{\ell'-1}\,\sigma_{k}\,
\Big((-1)^{(|\bt^0|)}\,h(\bt^1,z)\Big)^{\delta_{\ell<1}\,\delta_{k>0}}\ 
(k+1)^{\delta_{k,\ell'}}
\times\\[2mm]
&\times 
 \psi_\ell(z;\bt)\ \phi_k(z;\bt),
\quad \ell<k\,,
\end{split}
\end{equation}
\begin{equation}\label{mu-spb}
\begin{split}
\mu^\ell_\ell(z;\bt)&=
\psi_\ell(z;\bt)\ \phi_\ell(z;\bt)
 =\begin{cases}
f(\bt^\ell,z)\,f(z,\bt^{\ell-1})\,,&2\leq\ell\leq n\,,\\
f(\bt^1,z)\,f_0(z,\bt^{0})\,,&\ell=1\,,\\
f_0(\bt^{0},z)\,
f(z_{1},\bt^{1})\,,&\ell=0\,,\\
f(\bt^{-\ell},z_{-\ell})\,
f(z_{-\ell+1},\bt^{-\ell+1})\,,&-n+1\leq\ell\leq -1\,.\\
\end{cases}
\end{split}
\end{equation}
Recall that according to the definition \r{g-h-f-gen} the function 
$f_0(u,v)=\frac{u-v+2\,c}{u-v}$.

\item For $\ggo=\mathfrak{o}_{2n}$:
\begin{equation}\label{mu-o2na}
\begin{split}
\mu^k_\ell(z;\bt)&=\sigma_{\ell'-k}\ 
(-1)^{(|\bt^0|+|\bt^1|)\delta_{\ell<2}\,\delta_{k>-1}}\ 
(k-1)^{\delta_{k,\ell'}}
\ \psi_\ell(z;\bt)\ \phi_k(z;\bt),
\quad \ell<k\,,
\end{split}
\end{equation}
\begin{equation}\label{mu-o2nb}
\begin{split}
\mu^\ell_\ell(z;\bt)&=
(-1)^{(|\bt^0|+|\bt^1|)(\delta_{\ell,0}+\delta_{\ell,1})}\ 
\psi_\ell(z;\bt)\ \phi_\ell(z;\bt)=\\
&=\begin{cases}
f(\bt^\ell,z)\,f(z,\bt^{\ell-1})\,,&3\leq\ell\leq n\,,\\
f(\bt^2,z)\,f(z,\bt^{0})\,f(z,\bt^{1})\,,&\ell=2\,,\\
f(\bt^1,z)\,f(z,\bt^{0})\,,&\ell=1\,,\\
f(\bt^0,z)\,f(z,\bt^{1})\,,&\ell=0\,,\\
f(z_2,\bt^2)\,f(\bt^1,z)\,f(\bt^0,z)\,,&\ell=-1\,,\\
f(\bt^{-\ell},z_{-\ell})\,
f(z_{-\ell+1},\bt^{-\ell+1})\,,&-n+1\leq\ell\leq -2\,.
\end{cases}
\end{split}
\end{equation}

\end{subequations}
\end{itemize}

\section{Description of the subsets in proposition~\ref{diag-act}
\label{AppA}}

In this appendix we describe explicitly all the subsets in the partition of the 
extended sets $\bw^s$ which yield a non-zero contribution 
 in the action of the diagonal monodromy entries 
$T_{i,i}(z)$ on the off-shell Bethe vectors $\BB(\bt)$. 
We present this 
description  for  algebras $\ggo\neq\gln$. 
The description for $\gln$ has already been presented  
 in proposition~\ref{diag-act-gln}, 
formulas \r{bea1}.

For all three algebras $\ggo=\mathfrak{o}_{2n+1}$, $\mathfrak{sp}_{2n}$, 
and $\mathfrak{o}_{2n}$ we first formulate the following lemma. 

\begin{lemma}\label{prec-van}
Let $q\in\{{\so},{\st},{\sth}\}$ denote the index labeling 
 the subset $\bw^s_q$. We equip the set $\{{\so},{\st},{\sth}\}$ with the natural ordering 
${\so}\scriptstyle{\prec}{\st}$,  ${\so}\scriptstyle{\prec}{\sth}$,  and 
${\st}\scriptstyle{\prec}{\sth}$. 
The partitions of the sets 
$\bw^s\dashv \{\bw^s_{\so},\bw^s_{\st},\bw^s_{\sth}\}$ such that $\{z\}\in\bw^a_{q}$ and 
$\{z_b\}\in\bw^b_{q'}$ with $q\prec q'$ and 
$a,b=0,1,\ldots,n-1$ have a zero contribution to the action formula \r{T-act-all}.
\end{lemma}
\proof 
We provide the proof of this lemma in the case $\ggo=\mathfrak{o}_{2n+1}$. 
The two other  cases  $\ggo=\mathfrak{sp}_{2n}$ and 
$\ggo=\mathfrak{o}_{2n}$ can be proved analogously. 

Assume that $z\in\bw^a_q$ for some $a=1,\ldots,n-1$ and 
$q=\{{\so},{\st}\}$. The explicit form of the function 
$\Om(\bw_{\so},\bw_{\st},\bw_{\sth})$ \r{Om-full} yields 
that $z\not\in\bw^{a-1}_{q'}$ for $q\prec q'$ since otherwise
 the corresponding term in the sum over partitions 
vanishes due to $g(z,z)^{-1}=0$. This means that  
$z\in\bw^{a-1}_{q}$ and by induction over $a$ we get 
that $z\in\bw^0_q$. 

Assuming now that $z_b\in\bw^b_{q'}$ 
for some $b=1,\ldots,n-1$ and 
$q'=\{{\st},{\sth}\}$. Again, the explicit form of the function 
$\Om(\bw_{\so},\bw_{\st},\bw_{\sth})$ \r{Om-full} yields 
that $z_b\not\in\bw^{b-1}_{q}$ for $q\prec q'$ since otherwise  the corresponding term in the sum over partitions 
vanishes due to $h(z_b,z_{b-1})=0$. This means that  
$z_b\in\bw^{b-1}_{q'}$ and by induction over $b$  that $z_0\in\bw^0_{q'}$. 

But the factor $f_0(\bw^0_q,\bw^0_{q'})$ for $q\prec q'$ 
in the function $\Om(\bw_{\so},\bw_{\st},\bw_{\sth})$ \r{Om-full}
will vanish because $f_0(z,z_0)=0$. This proves that 
our assumptions that $z\in\bw^a_q$ and $z_b\in\bw^b_{q'}$
for $q\prec q'$ leads to a zero contribution in the sum over partitions. 
\qed

In the lemmas~\ref{wanted} and \ref{lem-all} below 
we describe the subsets $\bw^s_{\so}$,
$\bw^s_{\st}$, and $\bw^s_{\sth}$  
which yield non-zero contributions in the action of the 
diagonal monodromy entries $T_{i,i}(z)$ to off-shell Bethe vectors $\BB(\bt)$
in a generic $\ggo$-invariant integrable model.
Using the parameter $\bvphi_\ggo$ \eqref{vphi},
the description of such subsets can be done uniformly for all $\ggo\not=\gln$. 
Recall that the parameter $\bvphi_\ggo$ is such that
$\bvphi_{\ggb}=0$, $\bvphi_{\gsp}=1$, and 
$\bvphi_{\ggd}=2$.

\begin{lemma}\label{lem-all} Let $\ell$ and $k$ be integers from 
the set $\Ig$ such that $\ell\leq k$. 

\textbf{1. Fix a positive integer $i$ such that $\bvphi_\ggo\leq i\leq n$.}
In the sum over partitions 
$\bw^s\dashv\{\bw^s_{\so},\bw^s_{\st},\bw^s_{\sth}\}$ 
in \r{T-act-all} such that cardinalities of the subsets are 
given by \r{part-all},
the partitions which correspond to non-zero terms are labeled 
by a pair of integers $(\ell,k)$ such that  
$n'\leq\ell\leq i\leq k\leq n$.
They are given by the following list of formulas.

\begin{subequations}\label{Dig1}

\begin{itemize}

\item For $\bvphi_\ggo\leq \ell\leq i\leq k\leq n$
\begin{equation}\label{bec11a}
\begin{array}{ccccccc}
\bw^s_{\so}=\{z\},
&&\bw^s_{\st}=\bt^s,
&&\bw^s_{\sth}=\vn,
&&0\leq s<\bvphi_\ggo\,,\\[2mm]
\bw^s_{\so}=\{z_s,z\},
&&\bw^s_{\st}=\bt^s,
&&\bw^s_{\sth}=\vn,
&&\bvphi_\ggo\leq s\leq\ell-1\,,\\[2mm]
\bw^s_{\so}=\{z_s,\crd{\bt^s_{\so}}{1}\},
&&\bw^s_{\st}=\{z,\crd{\bt^s_{\st}}{m_s-1}\},
&&\bw^s_{\sth}=\vn,
&&\ell\leq s\leq i-1\,,\\[4mm]
\bw^s_{\so}=\{z_s\},
&&\bw^s_{\st}=\{z,\crd{\bt^s_{\st}}{m_s-1}\},
&&\bw^s_{\sth}=\{\crd{\bt^s_{\sth}}{1}\},
&&i\leq s\leq k-1\,,\\[4mm]
\bw^s_{\so}=\{z_s\},
&&\bw^s_{\st}=\bt^s,
&&\bw^s_{\sth}=\{z\},
&&k\leq s\leq n-1\,.
\end{array}
\end{equation}

\item For $\bvphi_\ggo\leq\ell'\leq i\leq k\leq n$
\begin{equation}\label{bec11b}
\begin{array}{ccccccc}
\bw^s_{\so}=\{\crd{\bt^s_{\so}}{1}\}\,,
&&\bw^s_{\st}=\{z,\crd{\bt^s_{\st}}{m_s-1}\},
&&\bw^s_{\sth}=\vn,
&&0\leq s<\bvphi_\ggo\,,\\[4mm]
\bw^s_{\so}=\{\crd{\bt^s_{\so}}{2}\}\,,
&&\bw^s_{\st}=\{z_s,z,\crd{\bt^s_{\st}}{m_s-2}\},
&&\bw^s_{\sth}=\vn,
&&\bvphi_\ggo\leq s\leq \ell'-1\,,\\[4mm]
\bw^s_{\so}=\{z_s,\crd{\bt^s_{\so}}{1}\},
&&\bw^s_{\st}=\{z,\crd{\bt^s_{\st}}{m_s-1}\},
&&\bw^s_{\sth}=\vn,
&&\ell'\leq s\leq i-1\,,\\[4mm]
\bw^s_{\so}=\{z_s\},
&&\bw^s_{\st}=\{z,\crd{\bt^s_{\st}}{m_s-1}\},
&&\bw^s_{\sth}=\{\crd{\bt^s_{\sth}}{1}\},
&&i\leq s\leq k-1\,,\\[4mm]
\bw^s_{\so}=\{z_s\},
&&\bw^s_{\st}=\bt^s,
&&\bw^s_{\sth}=\{z\},
&&k\leq s\leq n-1\,.
\end{array}
\end{equation}

 \item For $\bvphi_\ggo\leq i\leq \ell'\leq k\leq n$
\begin{equation}\label{bec11bb}
\begin{array}{ccccccc}
\bw^s_{\so}=\{\crd{\bt^s_{\so}}{1}\}\,,
&&\bw^s_{\st}=\{z,\crd{\bt^s_{\st}}{m_s-1}\},
&&\bw^s_{\sth}=\vn,
&&0\leq s<\bvphi_\ggo\,,\\[4mm]
\bw^s_{\so}=\{\crd{\bt^s_{\so}}{2}\}\,,
&&\bw^s_{\st}=\{z_s,z,\crd{\bt^s_{\st}}{m_s-2}\},
&&\bw^s_{\sth}=\vn,
&&\bvphi_\ggo\leq s\leq i-1\,,\\[4mm]
\bw^s_{\so}=\{\crd{\bt^s_{\so}}{1}\},
&&\bw^s_{\st}=\{z_s,z,\crd{\bt^s_{\st}}{m_s-2}\},
&&\bw^s_{\sth}=\{\crd{\bt^s_{\sth}}{1}\},
&&i\leq s\leq \ell'-1\,,\\[4mm]
\bw^s_{\so}=\{z_s\},
&&\bw^s_{\st}=\{z,\crd{\bt^s_{\st}}{m_s-1}\},
&&\bw^s_{\sth}=\{\crd{\bt^s_{\sth}}{1}\},
&&\ell'\leq s\leq k-1\,,\\[4mm]
\bw^s_{\so}=\{z_s\},
&&\bw^s_{\st}=\bt^s,
&&\bw^s_{\sth}=\{z\},
&&k\leq s\leq n-1\,,
\end{array}
\end{equation}

\item For $\bvphi_\ggo\leq i\leq k\leq \ell' \leq n$
\begin{equation}\label{bec11c}
\begin{array}{ccccccc}
\bw^s_{\so}=\{\crd{\bt^s_{\so}}{1}\},
&&\bw^s_{\st}=\{z,\crd{\bt^s_{\st}}{m_s-1}\},
&&\bw^s_{\sth}=\vn,
&&0\leq s<\bvphi_\ggo \,,\\[4mm]
\bw^s_{\so}=\{\crd{\bt^s_{\so}}{2}\},
&&\bw^s_{\st}=\{z_s,z,\crd{\bt^s_{\st}}{m_s-2}\},
&&\bw^s_{\sth}=\vn,
&&\bvphi_\ggo\leq s\leq i-1\,,\\[4mm]
\bw^s_{\so}=\{\crd{\bt^s_{\so}}{1}\},
&&\bw^s_{\st}=\{z_s,z,\crd{\bt^s_{\st}}{m_s-2}\},
&&\bw^s_{\sth}=\{\crd{\bt^s_{\sth}}{1}\},
&&i\leq s\leq k-1\,,\\[4mm]
\bw^s_{\so}=\{\crd{\bt^s_{\so}}{1}\}
&&\bw^s_{\st}=\{z_s,\crd{\bt^s_{\st}}{m_s-1}\}
&&\bw^s_{\sth}=\{z\},
&&k\leq s\leq \ell'-1\,,\\[4mm]
\bw^s_{\so}=\{z_s\},
&&\bw^s_{\st}=\bt^s,
&&\bw^s_{\sth}=\{z\},
&&\ell'\leq s\leq n-1\,.
\end{array}
\end{equation}

\end{itemize}
\end{subequations}

\textbf{2. Fix a non-positive integer $i$ such that $\bvphi_\ggo\leq i'\leq n$.} 
In \r{T-act-all} with $j=i$, the partitions which have non-zero contributions
 are given by the same equalities 
\r{Dig1} with the replacements 
\begin{equation}\label{replace}
i\to i',\qquad \ell\leftrightarrow k',\qquad
{\so}\leftrightarrow {\sth},\qquad \{z\}\leftrightarrow \{z_s\}. 
\end{equation}

For the algebras $\ggo=\ggb$ and $\ggo=\gsp$ 
the subsets given by the formulas 
\r{Dig1} together with ones 
obtained by the replacement \r{replace} describe 
all the subsets which yield non-zero contribution to the action of diagonal 
monodromy entries on off-shell Bethe vectors.
Note that the first lines in the formulas \r{Dig1} are absent for the algebra 
$\ggo=\ggb$ since $\bvphi_{\ggb}=0$. 

For the algebra $\ggo=\ggd$, $\bvphi_{\ggd}=2$ and we 
need to consider two additional cases. 

\smallskip

\textbf{
 3. For $\ggo=\ggd$ and $i,k,\ell$ such that $2\leq i\leq k\leq n$
 and  $0\leq \ell\leq 1$ }

\begin{equation}\label{lbed32}
\begin{array}{ccccccc}
\bw^s_{\so}=\{z\},
&&\bw^s_{\st}=\bt^s,
&&\bw^s_{\sth}=\vn,
&&s=\ell'\,,\\[2mm]
\bw^s_{\so}=\{\crd{\bt^s_{\so}}{1}\},
&&\bw^s_{\st}=\{z,\crd{\bt^s_{\st}}{m_s-1}\},
&&\bw^s_{\sth}=\vn,
&&s=\ell\,,\\[2mm]
\bw^s_{\so}=\{z_s,\crd{\bt^s_{\so}}{1}\},
&&\bw^s_{\st}=\{z,\crd{\bt^s_{\st}}{m_s-1}\},
&&\bw^s_{\sth}=\vn,
&&2\leq s\leq i-1\,,\\[4mm]
\bw^s_{\so}=\{z_s\},
&&\bw^s_{\st}=\{z,\crd{\bt^s_{\st}}{m_s-1}\},
&&\bw^s_{\sth}=\{\crd{\bt^s_{\sth}}{1}\},
&&i\leq s\leq k-1\,,\\[4mm]
\bw^s_{\so}=\{z_s\},
&&\bw^s_{\st}=\bt^s,
&&\bw^s_{\sth}=\{z\},
&&k\leq s\leq n-1\,.
\end{array}
\end{equation}

\bigskip

\textbf{4. For $\ggo=\ggd$ and $i=1$ the subsets are}

\begin{subequations}\label{Die1}

\begin{itemize} 

\item for $\ell=k=1$ 
\begin{equation}\label{want2}
\begin{array}{ccccccc}
\bw^0_{\so}=\{z\},
&&\bw^0_{\st}=\bt^0,
&&\bw^0_{\sth}=\vn,
&&s=0\,,\\[2mm]
\bw^1_{\so}=\vn,
&&\bw^1_{\st}=\bt^1,
&&\bw^1_{\sth}=\{z\},
&&s=1\,,\\[2mm]
\bw^s_{\so}=\{z_s\},
&&\bw^s_{\st}=\bt^s,
&&\bw^s_{\sth}=\{z\},
&&2\leq s\leq n-1\,,
\end{array}
\end{equation}

\item for $\ell=1$ and $2\leq k\leq n$
\begin{equation}\label{lbed42}
\begin{array}{ccccccc}
\bw^0_{\so}=\{z\},
&&\bw^0_{\st}=\bt^0,
&&\bw^0_{\sth}=\vn,
&&s=0\,,\\[2mm]
\bw^1_{\so}=\vn,
&&\bw^1_{\st}=\{z,\bt^1_{\st}\},
&&\bw^1_{\sth}=\bt^1_{\sth},
&&s=1\,,\\[2mm]
\bw^s_{\so}=\{z_s\},
&&\bw^s_{\st}=\{z,\bt^s_{\st}\},
&&\bw^s_{\sth}=\bt^s_{\sth},
&&2\leq s\leq k-1\,,\\[2mm]
\bw^s_{\so}=\{z_s\},
&&\bw^s_{\st}=\bt^s,
&&\bw^s_{\sth}=\{z\},
&&k\leq s\leq n-1\,,
\end{array}
\end{equation}

\item for $2\leq\ell'\leq n$ and $k=1$ 
\begin{equation}\label{lbed44}
\begin{array}{ccccccc}
\bw^0_{\so}=\bt^0_{\so}\,,
&&\bw^0_{\st}=\{z,\bt^0_{\st}\},
&&\bw^0_{\sth}=\vn,
&&s=0\,,\\[2mm]
\bw^1_{\so}=\vn,
&&\bw^1_{\st}=\bt^1,
&&\bw^1_{\sth}=\{z\},
&&s=1\,,\\[2mm]
\bw^s_{\so}=\bt^s_{\so}\,,
&&\bw^s_{\st}=\{z_s,\bt^s_{\st}\},
&&\bw^s_{\sth}=\{z\},
&&2\leq s\leq \ell'-1\,,\\[2mm]
\bw^s_{\so}=\{z_s\},
&&\bw^s_{\st}=\bt^s,
&&\bw^s_{\sth}=\{z\},
&&\ell'\leq s\leq n-1\,,
\end{array}
\end{equation}

\item for $2\leq \ell'\leq k\leq n$
\begin{equation}\label{lbed46}
\begin{array}{ccccccc}
\bw^0_{\so}=\bt^0_{\so},
&&\bw^0_{\st}=\{z,\bt^0_{\st}\},
&&\bw^0_{\sth}=\vn,
&&s=0\,,\\[2mm]
\bw^1_{\so}=\vn,
&&\bw^1_{\st}=\{z,\bt^1_{\st}\},
&&\bw^1_{\sth}=\bt^1_{\sth},
&&s=1\,,\\[2mm]
\bw^s_{\so}=\bt^s_{\so},
&&\bw^s_{\st}=\{z_s,z,\bt^s_{\st}\},
&&\bw^s_{\sth}=\bt^s_{\sth},
&&2\leq s\leq \ell'-1\,,\\[2mm]
\bw^s_{\so}=\{z_s\},
&&\bw^s_{\st}=\{z,\bt^s_{\st}\},
&&\bw^s_{\sth}=\bt^s_{\sth},
&&\ell'\leq s\leq k-1\,,\\[2mm]
\bw^s_{\so}=\{z_s\},
&&\bw^s_{\st}=\bt^s,
&&\bw^s_{\sth}=\{z\},
&&k\leq s\leq n-1\,,
\end{array}
\end{equation}

\item for $2\leq k\leq \ell' \leq n$
\begin{equation}\label{lbed47}
\begin{array}{ccccccc}
\bw^0_{\so}=\bt^0_{\so},
&&\bw^0_{\st}=\{z,\bt^0_{\st}\},
&&\bw^0_{\sth}=\vn,
&&s=0 \,,\\[2mm]
\bw^1_{\so}=\vn,
&&\bw^1_{\st}=\{z,\bt^1_{\st}\},
&&\bw^1_{\sth}=\bt^1_{\sth},
&&s=1 \,,\\[2mm]
\bw^s_{\so}=\bt^s_{\so},
&&\bw^s_{\st}=\{z_s,z,\bt^s_{\st}\},
&&\bw^s_{\sth}=\bt^s_{\sth},
&&2\leq s\leq k-1\,,\\[2mm]
\bw^s_{\so}=\bt^s_{\so},
&&\bw^s_{\st}=\{z_s,\bt^s_{\st}\},
&&\bw^s_{\sth}=\{z\},
&&k\leq s\leq \ell'-1\,,\\[2mm]
\bw^s_{\so}=\{z_s\},
&&\bw^s_{\st}=\bt^s,
&&\bw^s_{\sth}=\{z\},
&&\ell'\leq s\leq n-1\,.
\end{array}
\end{equation}

\end{itemize}
\end{subequations}
The cardinalities given by the formulas \r{lbed32} and \r{Die1} together with the 
ones obtained from the replacement \r{replace} 
describe the remaining cases for the subsets in the partitions 
$\bw^s\dashv\{\bw^s_\so,\bw^s_\st,\bw^s_\sth\}$ 
which yield non-zero contribution to the action \r{d-ac1} of diagonal 
monodromy entries in $\ggd$-invariant integrable models.

\end{lemma}

\proof
We provide the proof of this lemma~\ref{lem-all} 
 for the case $\ggo=\ggb$ and $1\leq i\leq n$. 
The proofs for other cases is similar. 

The first step is to give the description of the subsets $\bw_\sth$ given by 
the third columns in the formulas \r{Dig1}. 
Since  the index $i$ is strictly positive it means that, 
according to the description  
\r{partB}, the cardinality $|\bw^s_{\sth}|$ 
can be either 0 for $0\leq s\leq i-1$ or 1 for $i\leq s\leq n-1$. 
Consider the latter case when $|\bw^s_{\sth}|=1$. Then, according 
to the lemma~\ref{prec-van} the subset $\bw^s_{\sth}$ can be
either $\{z\}$ or $\bt^s_{\sth}$. Fix a positive integer $a$ 
such that $i\leq a\leq n-1$ and assume that 
$\bw^a_{\sth}=\bt^a_{\sth}$. 
 Then the structure of the function  
$\Om(\bw_{\so},\bw_{\st},\bw_{\sth})$ \r{Om-full} fixes 
that $\bw^s_{\sth}=\bt^s_{\sth}$ for all $s$ such that $i\leq s\leq a-1$. 
Indeed, our assumption $\bw^a_{\sth}=\bt^a_{\sth}$
implies that $z\in\bw^a_{{\so},{\st}}$. But then, the factor 
$g(\bw^a_{{\so},{\st}},\bw^{a-1}_{\sth})^{-1}$ fixes that 
$\bw^{a-1}_{\sth}=\bt^{a-1}_{\sth}$ since otherwise 
$\bw^{a-1}_{\sth}=\{z\}$ and this term vanishes in the sum over 
partitions  due to $g(z,z)^{-1}=0$. 
These considerations show that there exist a positive integer $k$,
$i\leq k\leq n$ which labels the partitions and
such that the subsets  $\bw^s_{\sth}$ are $\{z\}$ for $k\leq s\leq n-1$ and 
$\bt^s_{\sth}$ (of cardinality 1) for $i\leq s\leq k-1$. 
This proves the third column in the formulas \r{Dig1} 
of the lemma~\ref{lem-all} and fixes the positive integer $k$.

According to the first formula in \r{partB} the cardinality of 
the subset $\bw^s_\so$ for $1\leq i\leq n$ are equal to 
2 for $0\leq s\leq i-1$ and to 1 for $i\leq s\leq n-1$. 
We consider these two cases separately. 

First, we consider 
the case when $|\bw^s_\so|=1$, namely the values of 
$s$ such that $i\leq s\leq n-1$. This case corresponds to 
last two lines in \r{bec11a}, \r{bec11b} and last three lines in 
\r{bec11bb} and \r{bec11c}.

Fix a positive integer $b$ such that $i\leq b\leq n-1$. Then according to 
\r{partB} the cardinality $|\bw^b_{\so}|=1$ and the subset 
$\bw^b_{\so}$ can be either $\{z_b\}$ or $\bt^b_{\so}$. The case 
$\bw^b_{\so}=\{z\}$ is excluded by the lemma~\ref{prec-van}. 
Let us assume that $\bw^b_{\so}=\bt^b_{\so}$. Then the structure 
of the function $\Om(\bw_{\so},\bw_{\st},\bw_{\sth})$ \r{Om-full}
fixes that $\bw^s_{\so}=\bt^s_{\so}$ for all $s$ such that $i\leq s\leq b-1$.
Indeed, an assumption  $\bw^b_{\so}=\bt^b_{\so}$ means that 
$\bw^b_{{\st},{\sth}}=\{z_b,z,\bt^b_{{\st},{\sth}}\}$ and 
the factor $h(\bw^b_{{\st},{\sth}},\bw^{b-1}_{\so})$ fixes that 
$\bw^{b-1}_{\so}=\bt^{b-1}_{\so}$ since otherwise 
the term in the sum over partition such that 
$\bw^{b-1}_{\so}=\{z_{b-1}\}$ vanishes due to $h(z_b,z_{b-1})=0$. 
In turn, it signifies that there exist an positive integer $\ell_1$,
$i\leq\ell_1\leq n$ which labels the partitions  and such that 
$\bw^{s}_{\so}=\{z_{s}\}$ for $\ell_1\leq s\leq n-1$ and 
$\bw^{s}_{\so}=\bt^{s}_{\so}$ for $i\leq s\leq \ell_1-1$.

Now we have to consider the different cases of interrelation 
between the positive integers $\ell_1$, $i$ and $k$. 
\begin{itemize}
\item Let $i\leq k\leq \ell_1\leq n$ and $i\leq \ell_1\leq k\leq n$. 
Then according to the analysis above 
$\bw^s=\bt^s_\so$ for $i\leq s\leq \ell_1$ and $\bw^s=\{z_s\}$
for $\ell_1\leq s\leq n-1$. Combining this result with the already proven 
description of the subsets $\bw^s_\sth$ we obtain last three lines in 
\r{bec11bb} and \r{bec11c} 
where the positive integer $\ell_1$ is replaced by $\ell'$ such 
that $-k\leq\ell\leq -i$ in the first case and $-n\leq\ell\leq -k$ in the second one.

\item Let $1\leq \ell_1\leq i\bleu{\leq k\leq n}$. Then in the last two lines of both 
formulas \r{bec11a} and  \r{bec11b} one gets that the 
subsets $\bw^s_\so$  are equal to $\{z_s\}$. Together with the already shown
description of the sets $\bw^s_\sth$ for $i\leq s\leq n-1$ we proved 
the last two lines in \r{bec11a} and  \r{bec11b} which are the same and 
do not depend on $\ell_1$. 
\end{itemize}

Now we repeat similar arguments for the cardinality 2 subsets 
$\bw^s_\so$ for $0\leq s\leq i-1$. Recall again that 
according to \r{partB} the subset $\bw^d_{\sth}$ 
is empty for these values of $s$.

Fix again a non-negative integer $d$ such that $0\leq d\leq i-1$. 
For these values of $d$ the subset 
$\bw^d_{\so}$ has a cardinality 2 and can be either 
$\{z_d,\bt^d_{\so}\}$ with $|\bt^d_{\so}|=1$ or $\{z_d,z\}$ or 
$\bt^d_{\so}$ with $|\bt^d_{\so}|=2$. 
The case when $\bw^d_{\so}=\{z,\bt^d_{\so}\}$ is excluded by 
the lemma~\ref{prec-van}.

Let us assume that 
 $\bw^d_{\so}=\{z_d,z\}$. Then we can prove that 
$\bw^s_{\so}=\{z_s,z\}$ for all $0\leq s\leq d-1$. Indeed, 
according to our assumption 
the factor $g(\bw^d_{\so},\bw^{d-1}_{\st})^{-1}$
fixes that $z\not\in \bw^{d-1}_{\st}$ otherwise this term in the sum over 
partitions vanishes due to $g(z,z)^{-1}=0$. This means 
that $z\in\bw^{d-1}_{\so}$ and the lemma~\ref{prec-van} 
provides that $z_{d-1}$ also belongs to the subset $\bw^{d-1}_{\so}$. 

On the other hand if $\bw^{d}_{\so}=\{z_d,\bt^d_{\so}\}$ which means that
$\bw^{d}_{\st}=\{z,\bt^d_{\st}\}$ then we can show that 
$\bw^{s}_{\so}=\{z_s,\bt^s_{\so}\}$ and 
$\bw^{s}_{\st}=\{z,\bt^s_{\st}\}$ for $d+1\leq s\leq i-1$. 
Indeed the factor $h(\bw^{d+1}_{\st},\bw^d_{\so})$ fixes that 
$z_{d+1}\not\in\bw^{d+1}_{\st}$ and the factor 
$g(\bw^{d+1}_{\so},\bw^d_{\st})^{-1}$ fixes that 
$z\not\in\bw^{d+1}_{\so}$ which proves the assertion. 

These considerations 
 mean that there exist a non-negative integer $\ell$
such that $0\leq\ell\leq i$, 
$\bw^{s}_{\so}=\{z_s,z\}$, $\bw^s_{\st}=\bt^s$ for $0\leq s\leq \ell-1$
and 
$\bw^{s}_{\so}=\{z_s,\bt^s_{\so}\}$, 
$\bw^s_{\st}=\{z,\bt^s_{\st}\}$ for $\ell\leq s\leq i-1$. 
This proves second and third lines in \r{bec11a}. Recall that the first lines 
in \r{Dig1} are absent.  

The final case corresponds to $\bw^d_\so=\bt^d_\so$ (of cardinality 2) for some 
$0\leq d\leq i-1$. 
Again the structure of the 
function $\Om(\bw_{\so},\bw_{\st},\bw_{\sth})$ \r{Om-full}
fixes that $\bw^s_{\so}=\bt^s_{\so}$ for all $s$ such that 
$0\leq s\leq d-1$. Indeed, the assumption $\bw^d_{\so}=\bt^d_{\so}$
means that $\bw^d_{\st}=\{z_d,z,\bt^d_{\st}\}$. Then 
the factor $h(\bw^{d}_{\st},\bw^{d-1}_{\so})$ fixes that 
$z_{d-1}\not\in\bw^{d-1}_{\so}$ which implies that 
$\bw^{d-1}_{\so}=\{\bt^{d-1}_{\so}\}$ again with  
$|\bt^{d-1}_{\so}|=2$. Otherwise 
if $z_{d-1}\in\bw^{d-1}_{\so}$ then this term  
vanishes due to $h(z_d,z_{d-1})=0$.

These considerations 
show that there exist a non-negative integer $\ell_2$, $0\leq\ell_2\leq i$,
labeling the partitions and such that 
$\bw^s_{\so}=\bt^s_{\so}$,
$\bw^s_{\st}=\{z_s,z,\bt^s_{\st}\}$ for  $0\leq s\leq \ell_2-1$
with $|\bt^s_{\so}|=2$ and 
$\bw^s_{\so}=\{z_s,\bt^s_{\so}\}$,
$\bw^s_{\st}=\{z,\bt^s_{\st}\}$ for  $\ell_2\leq s\leq i-1$
with $|\bt^s_{\so}|=1$. 

If $\ell_2=i$ then $\bw^s_{\so}=\bt^s_{\so}$,
$\bw^s_{\st}=\{z_s,z,\bt^s_{\st}\}$ for  $0\leq s\leq i-1$
with $|\bt^s_{\so}|=2$: it 
yields to second lines in \r{bec11bb} and \r{bec11c}. 

For $0\leq \ell_2 <i$, we can redefine $\ell_2=\ell'$ with $-i<\ell\leq 0$. 
It proves second and third lines in \r{bec11b}. 

Summarizing we conclude that 
in the case $1\leq i\leq n$ the 
subsets $\bw^s_\so$, $\bw^s_\st$, and $\bw^s_\sth$
which yields non-zero contribution in the action of the 
diagonal monodromy entries $T_{i,i}(z)$ on the off-shell Bethe vectors 
$\BB(\bt)$ are labeled by a pair of integers $\ell$ and $k$ such that 
$n'\leq\ell\leq i\leq k\leq n$ and presented by the formulas 
\r{Dig1}. The cases of other $i$ can be  considered similarly. 
\qed

Once we have determined the partitions leading to non-zero contributions, one can select the ones 
associated to the \textsl{wanted} terms. Indeed, these wanted terms just correspond to the partitions such that 
$\ell=k=i$ and $\bar t_{\st}=\bar t$.
A careful examination of the partitions listed in lemma \ref{lem-all} leads to the following lemma.

\begin{lemma}\label{wanted}
The subsets $\bw^s_\so$, $\bw^s_\st$, and $\bw^s_\sth$ which yield 
nonzero contributions to the diagonal monodromy action \r{T-act-all} 
and corresponding to wanted terms for $\ggo\not=\gln$
and $\bvphi_\ggo\leq i\leq n$ are given by the equalities 
\begin{equation}\label{want1}
\begin{array}{ccccccc}
\bw^s_{\so}=\{z\},
&&\bw^s_{\st}=\bt^s,
&&\bw^s_{\sth}=\vn,
&&0\leq s<\bvphi_\ggo\,,\\[2mm]
\bw^s_{\so}=\{z_s,z\},
&&\bw^s_{\st}=\bt^s,
&&\bw^s_{\sth}=\vn,
&&\bvphi_\ggo\leq s\leq i-1\,,\\[2mm]
\bw^s_{\so}=\{z_s\},
&&\bw^s_{\st}=\bt^s,
&&\bw^s_{\sth}=\{z\},
&&i\leq s\leq n-1
\end{array}
\end{equation}
and for  $i=1$ when $\ggo=\ggd$ by the equalities \r{want2}. 
The subsets which leads to the wanted terms for $\bvphi_\ggo\leq i'\leq n$
when $\ggo\not=\gln$  are given by the 
same equalities \r{want1} and for $i=0$ when $\ggo=\ggd$ by the equalities \r{want2} with the replacement 
\begin{equation}\label{rep-wan}
i\to i',\qquad 
{\so}\leftrightarrow {\sth},\qquad \{z\}\leftrightarrow \{z_s\}. 
\end{equation}
\end{lemma}

\paragraph{Cardinalities of the $\bar t$ subsets.} For all $\ggo$ the cardinalities of the 
subsets $\bt^s_{\so}$, $\bt^s_{\st}$,
 and $\bt^s_{\sth}$ in the action of diagonal monodromy entries 
 on the off-shell Bethe vectors 
 are given by the equalities

 \begin{equation}\label{C-eig-g}
\begin{split}
|\bt^s_{\so}|&=\begin{cases}
\Theta(i-\bvphi_\ggo)
\Theta(\ell'-\bvphi_\ggo)+
\Big(\Theta(i-\bvphi_\ggo)\,\delta_{s,\ell}
+\Theta(\ell'-\bvphi_\ggo)\,\delta_{s,i'}\Big)\,\delta_{\bvphi_\ggo,2}
\,,\quad 0\leq s< \bvphi_\ggo\,,\\
\Theta(i-\bvphi_\ggo)
\Big(\Theta(\ell'+s-\xi_\ggo)+\Theta(\ell'-s-1)\Big)\,,
\quad \bvphi_\ggo\leq s<|i|_\ggo,,\\
\Theta(\ell'-s-1)\Theta(N_\ggo-\xi_\ggo)\,,
\quad |i|_\ggo+\delta_{\bvphi_\ggo,2}\leq s\leq n-1\,,
\end{cases}\\
|\bt^s_{\sth}|&=\begin{cases}
\Theta(i'-\bvphi_\ggo)
\Theta(k-\bvphi_\ggo)+
\Big(\Theta(i'-\bvphi_\ggo)\,\delta_{s,k'}
+\Theta(k-\bvphi_\ggo)\,\delta_{s,i}\Big)\,\delta_{\bvphi_\ggo,2}
\,,\quad 0\leq s< \bvphi_\ggo\,,\\
\Theta(i'-\bvphi_\ggo)
\Big(\Theta(k+s-\xi_\ggo)+\Theta(k-s-1)\Big)\,,
\quad \bvphi_\ggo\leq s<|i|_\ggo,,\\
\Theta(k-s-1)\,,
\quad |i|_\ggo+\delta_{\bvphi_\ggo,2}\leq s\leq n-1\,.
\end{cases}
\end{split}
\end{equation}

Note that the formulas \r{C-eig-g} can be mapped to the formulas 
\r{oC-g} describing the cardinalities in the recurrence relations 
with following replacement in the latter formulas 
\begin{equation}\label{last-repl}
\ell\to i,\qquad k\to i,\qquad i\to \ell,\qquad j\to k\,.
\end{equation}
Note that $i=j$ in \r{C-eig-g} corresponds to $k=\ell$ in \r{oC-g}.

\section*{Data Availability Statement}

No datasets were generated or analyzed during the current study. Therefore, data sharing is not applicable to this article.

\section*{Conflict of Interest}
The authors declare that there are no commercial or financial relationships that could be construed as a potential conflict of interest.

\end{document}